\documentclass[10pt,twoside]{article}
\usepackage{amsfonts}
\usepackage{amssymb}
\usepackage{amsmath}
\usepackage[dvips]{graphicx}
\usepackage[latin9]{inputenc}
\usepackage{pifont}
\usepackage{rotating}
\usepackage[T1]{fontenc}
\usepackage{geometry}
\usepackage{array}
\usepackage{units}
\usepackage{multirow}
\usepackage{esint}
\usepackage{footnote}

\begin{document}

\title{Solutions of\ Laplace's equation with simple boundary conditions, and
their applications for capacitors with multiple symmetries}
\author{Mayckol Morales\thanks{%
mjmoralesc@unal.edu.co}, Rodolfo A. Diaz\thanks{%
radiazs@unal.edu.co}, William J. Herrera\thanks{%
jherreraw@unal.edu.co}. \\
%EndAName
Departamento de Física. Universidad Nacional de Colombia. Bogotá, Colombia.}
\date{}
\maketitle

\begin{abstract}
We find solutions of Laplace's equation with \noindent specific boundary
conditions (in which such solutions take either the value zero or unity in
each surface) using a generic curvilinear system of coordinates. Such purely
geometrical solutions (that we shall call Basic Harmonic Functions BHF's)
are utilized to obtain a more general class of solutions for Laplace's
equation, in which the functions take arbitrary constant values on the
boundaries. On the other hand, the BHF's are also used to obtain the
capacitance of many electrostatic configurations of conductors. This method
of finding solutions of Laplace's equation and capacitances with multiple
symmetries is particularly simple, owing to the fact that the method of
separation of variables becomes much simpler under the boundary conditions
that lead to the BHF's. Examples of application in complex symmetries are
given. Then, configurations of succesive embedding of conductors are also
examined. In addition, expressions for electric fields between two
conductors and charge densities on their surfaces are obtained in terms of
generalized curvilinear coordinates. It worths remarking that it is
plausible to extrapolate the present method to other linear homogeneous
differential equations.

\end{abstract}

\section*{Introduction}

Solutions of Laplace's equation, usually called Harmonic Functions (HF's)\
are very important in many branches of Physics and Engineering such as
electrostatics, gravitation, hydrodynamics and thermodynamics\cite{Arfken,
Grif}. Therefore, considerable effort has been done in solving Laplace's
equation \cite{geomcapac3}-\cite{conformal1} in a variety of geometries and
boundary conditions. On the other hand, the capacitance of electrostatic
conductors is a very important quantity since capacitors are present in many
electric and electronic devices \cite{dasref}-\cite{cables}. Thus, many
studies of its general properties \cite{Grif}-\cite{cap08}, and calculations
of coefficients of capacitance for many geometric configurations \cite{cap02}%
-\cite{geomcapac3} have been carried out.

The first goal in this paper is to show a simple method to solve Laplace's
equation for configurations of volumes with certain symmetries, when the
HF's acquire values of either zero or unity in each surface that provides
the boundary conditions. The solutions of Laplace's equation under these
specific boundary conditions will be called Basic Harmonic Functions
(BHF's). We shall see that the form of these boundary conditions leads to
very easy solutions via separation of variables. Besides, the solutions will
be written in terms of a generic system of curvilinear coordinates that can
be adjusted for many symmetries. It is remarkable that the BHF's depends
exclusively on the geometry, and not on the physical problem involved.

The determination of the BHF's under a given geometry will be applied in two
scenarios: $\mathbf{(a)}$\ The generation of a more general class of
solutions in which the HF's take arbitrary constant values on the surfaces
that determine the volume. $\mathbf{(b)\ }$The calculation of the
coefficients of capacitance when the surfaces that form the volume are
covered with electrostatic conductors. Once again, both types of solutions
are given in a generic form in terms of appropriate generalized systems of
coordinates.

The paper is distributed as follows: In section \ref{Laplace}, we solve
Laplace's equation in generalized geometrical configurations to obtain the
BHF's, by using a generic system of curvilinear coordinates. As a matter of
illustration of the method, we obtain the BHF's for the cases of two
concentric spheres and two concentric cylinders. Then, from the BHF's
obtained in Sec. \ref{Laplace}\ and appealing to the linearity of Laplace's
equation, we construct in Sec. \ref{sec:equipotential case} a more general
class of solutions in which HF's acquire arbitrary constant values on the
surfaces. On the other hand, in section \ref{formulas of capacitance}, we
use the general expressions for the BHF's in generalized curvilinear
coordinates, in order to obtain the corresponding generalized formulas for
the coefficients of capacitance. The capacitance coefficients for the cases
of two concentric spheres and two concentric cylinders are obtained for
illustration. Then, section \ref{capaccalc} shows calculations of BHF's and
capacitances for more complex geometries in which our formulation acquires
all its power. Moreover, in appendix \ref{sec:field density}, we obtain the
electric field between two conductors as well as charge densities on their
surfaces in terms of curvilinear coordinates, while in appendix \ref%
{sec:embedding}, formulas for the coefficients of capacitance are extended
to the case in which we have a configuration of succesively embedded
conductors. Appendix \ref{ap:coordinate geom} provides some relations
between curvilinear coordinates and geometrical factors. Finally, section %
\ref{conclusions} contains our conclusions.

\section{Laplace's equation and systems of orthogonal curvilinear
coordinates \label{Laplace}}

Laplace's equation given by 
\begin{equation}
\nabla ^{2}f=0  \label{Laplaceeq}
\end{equation}%
has solutions that depend on the boundary conditions. It is desirable to use
a system of coordinates adjusted to the symmetry of the problem. Let us
consider a generic orthogonal coordinate system $\left( u,v,w\right) $, with
its corresponding scale factors $\left( h_{u},h_{v},h_{w}\right) $.
Furthermore, let us assume that the boundary consists of two closed surfaces 
$S_{1}$ and $S_{2}$ in which the coordinate $u$ takes constant values $u_{1}$
and $u_{2}\ $on $S_{1}$ and $S_{2}$ respectively. We shall suppose that one
of the surfaces contains the other, and we define as $V$ the volume
delimited between both closed surfaces. We intend to find a solution $%
f_{j}(u,v,w)$ of Laplace's equation in $V$,$\ $with the following boundary
conditions 
\begin{equation}
f_{j}\left( u_{j},v,w\right) =1\ ,\ \ f_{j}\left( u_{i},v,w\right) =0\ \ ;\ 
\text{with\ \ }i,j=1,2\ \text{ and\ \ }i\neq j  \label{eq:Condicion_Frontera}
\end{equation}%
where $i,j=1,2$ are indices that label the surfaces that determine the
boundary. Further, $u_{i}$ refers to the constant value of the coordinate $u$
on the surface $S_{i}$. As we already mentioned, we shall call Basic
Harmonic Functions (BHF's) to the solutions of Laplace's equation with the
boundary conditions defined in Eq. (\ref{eq:Condicion_Frontera}). In terms
of the $\left( u,v,w\right) $ coordinates, Laplace's equation for $f_{j}\ $%
becomes\ \cite{Arfken}\noindent 
\begin{equation}
\left[ \frac{\partial }{\partial u}\left( \frac{h_{v}h_{w}}{h_{u}}\frac{%
\partial f_{j}}{\partial u}\right) +\frac{\partial }{\partial v}\left( \frac{%
h_{w}h_{u}}{h_{v}}\frac{\partial f_{j}}{\partial v}\right) +\frac{\partial }{%
\partial w}\left( \frac{h_{u}h_{v}}{h_{w}}\frac{\partial f_{j}}{\partial w}%
\right) \right] =0  \label{eq:Laplaciano_Particular}
\end{equation}%
Assuming the following ansatz of separation 
\begin{equation}
f_{j}\left( u,v,w\right) =U_{j}(u)\Theta _{j}(v,w)\ \ ;\ \ j=1,2
\label{eq:Anzat1}
\end{equation}%
The boundary conditions Eq. (\ref{eq:Condicion_Frontera})\ yields 
\begin{eqnarray}
f_{j}\left( u_{j},v,w\right) =U_{j}\left( u_{j}\right) \Theta _{j}(v,w)
&=&1\ \ ;\ \ j=1,2  \label{boundary cond1} \\
f_{j}\left( u_{i},v,w\right) =U_{j}\left( u_{i}\right) \Theta _{j}(v,w)
&=&0\ \ ;\ \text{with\ \ }i,j=1,2\ \text{ and\ \ }i\neq j
\label{boundary cond2}
\end{eqnarray}

From now on, we shall assume that $i,j=1,2\ $and\ $i\neq j$ unless otherwise
stated. Since the coordinates $v$ and $w$ take any value on the surfaces, it
is necessary that $U_{j}\left( u_{i}\right) =0$, in order to satisfy the
boundary condition (\ref{boundary cond2}) and to obtain a non-trivial
solution. On the other hand, the boundary condition (\ref{boundary cond1})
says that 
\begin{equation}
\Theta _{j}\left( v,w\right) =\frac{1}{U_{j}\left( u_{j}\right) }\equiv A_{j}
\end{equation}%
where $A_{j}$ are two constants that only depend on $j$. Thus, the solution $%
f_{j}(u,v,w)$ in Eq. (\ref{eq:Anzat1})\ can then be rewritten as 
\begin{equation}
f_{j}(u,v,w)\equiv A_{j}U_{j}\left( u\right) \equiv f_{j}\left( u\right)
\label{fj(u)}
\end{equation}%
therefore the solution only depends on the coordinate $u$. The boundary
conditions (\ref{eq:Condicion_Frontera})\ become 
\begin{equation}
f_{j}\left( u_{i}\right) =0\ \ ;\ \ f_{j}\left( u_{j}\right) =1
\label{eq:Condiciones_Frontera}
\end{equation}%
from Eq. (\ref{fj(u)}) we get $\partial _{v}f_{j}=\partial _{w}f_{j}=0$ from
which Laplace's equation (\ref{eq:Laplaciano_Particular}) reads 
\begin{eqnarray}
\frac{\partial }{\partial u}\left( \frac{h_{v}h_{w}}{h_{u}}\frac{\partial
f_{j}\left( u\right) }{\partial u}\right) &=&0\ \ \Rightarrow
\label{eq:Laplaciano_Redox} \\
\frac{h_{v}h_{w}}{h_{u}}\frac{\partial f_{j}\left( u\right) }{\partial u}
&=&k_{j}\left( v,w\right)  \label{eq:Laplaciano_Redox2}
\end{eqnarray}%
note that there is not dependence on the coordinate $u\ $on the right hand
side of Eq. (\ref{eq:Laplaciano_Redox2}). Consequently, the $u-$dependence
given by $\partial _{u}f_{j}\left( u\right) $ on the left hand side of Eq. (%
\ref{eq:Laplaciano_Redox2}) must be cancelled by the remaining factor $%
h_{v}h_{w}/h_{u}$. It suggests that the dependence with $u$ on the term $%
h_{v}h_{w}/h_{u}$ should be factorized. Therefore, it is reasonable to
assume that 
\begin{equation}
\frac{h_{v}h_{w}}{h_{u}}=G\left( u\right) H\left( v,w\right)
\label{eq:Ansatz2}
\end{equation}

Note that the validity of this ansatz depends only on the coordinate system%
\footnote{%
In practice, the ansatz of separation (\ref{eq:Ansatz2}) holds in many
systems of curvilinear coordinates.}. Thus, neither $G\left( u\right) $ nor $%
H\left( v,w\right) $ could depend on $j$.\ Applying such an ansatz on Eq. (%
\ref{eq:Laplaciano_Redox}) we obtain

\begin{eqnarray}
H\left( v,w\right) \frac{\partial }{\partial u}\left[ G\left( u\right) \frac{%
\partial f_{j}\left( u\right) }{\partial u}\right] &=&0\ \ \Rightarrow \ \ 
\frac{d}{du}\left[ G\left( u\right) \frac{df_{j}\left( u\right) }{du}\right]
=0 \\
&\Rightarrow &\ G\left( u\right) \frac{df_{j}\left( u\right) }{du}=B_{j}
\end{eqnarray}%
where $B_{j}$ is another couple of constants for each $S_{j}$.\ We can solve
for$\ f_{j}\left( u\right) $ as follows 
\begin{eqnarray}
df_{j}\left( u\right) &=&\frac{B_{j}}{G\left( u\right) }du\ \ \Rightarrow \
\ f_{j}\left( u\right) =B_{j}\int \frac{du}{G\left( u\right) }+C_{j}  \notag
\\
f_{j}\left( u\right) &=&B_{j}Z\left( u\right) +C_{j}\ \ ;\ \ \frac{dZ\left(
u\right) }{du}\equiv \frac{1}{G\left( u\right) }  \label{eq:fj=00003DBjZ+Cj}
\end{eqnarray}%
once again $C_{j}$ is another couple of constants associated with each
surface. Combining (\ref{eq:fj=00003DBjZ+Cj}) with the conditions (\ref%
{eq:Condiciones_Frontera}) we have 
\begin{eqnarray}
f_{j}\left( u_{i}\right) &=&0=B_{j}Z\left( u_{i}\right) +C_{j}
\label{eq:Condicion_i_j} \\
f_{j}\left( u_{j}\right) &=&1=B_{j}Z\left( u_{j}\right) +C_{j}
\label{eq:Condcion_j_j}
\end{eqnarray}%
From (\ref{eq:Condicion_i_j}) it is obtained that 
\begin{equation}
C_{j}=-B_{j}Z\left( u_{i}\right)  \label{eq:Cj}
\end{equation}%
Substituting (\ref{eq:Cj}) in (\ref{eq:Condcion_j_j}) we obtain 
\begin{equation}
1=B_{j}Z\left( u_{j}\right) -B_{j}Z\left( u_{i}\right)
\end{equation}%
\begin{equation}
B_{j}=\frac{1}{\left[ Z\left( u_{j}\right) -Z\left( u_{i}\right) \right] }
\label{def Bj}
\end{equation}%
and substituting (\ref{eq:Cj}) and (\ref{def Bj}) in (\ref%
{eq:fj=00003DBjZ+Cj}) yields 
\begin{equation}
f_{j}\left( u\right) =\frac{Z\left( u\right) }{\left[ Z\left( u_{j}\right)
-Z\left( u_{i}\right) \right] }-B_{j}Z\left( u_{i}\right) =\frac{Z\left(
u\right) }{\left[ Z\left( u_{j}\right) -Z\left( u_{i}\right) \right] }-\frac{%
Z\left( u_{i}\right) }{\left[ Z\left( u_{j}\right) -Z\left( u_{i}\right) %
\right] }  \notag
\end{equation}

Summarizing, let us assume a geometric configuration with two closed
surfaces $S_{1}$ and $S_{2}$ in which one of the surfaces contains the
other. Both surfaces determine a volume $V$. Let us also assume that there
is a system of coordinates $\left( u,v,w\right) $, with scale factors $%
\left( h_{u},h_{v},h_{w}\right) $ such that the coordinate $u$ is constant
in each surface $S_{1}$ and $S_{2}$ (both constants $u_{1}$ and $u_{2}\ $%
should be different if we want a non-trivial solution). Under such
hypotheses, the solutions of Laplace's equation within the volume $V$ and
that satisfies the boundary conditions%
\begin{equation}
f_{j}\left( u_{j},v,w\right) =1\ ,\ \ f_{j}\left( u_{i},v,w\right) =0\ \ ;\ 
\text{with\ \ }i,j=1,2\ \text{ and\ \ }i\neq j
\label{eq:Condicion_Frontera2}
\end{equation}%
possess the following features: These solutions $f_{j}\left( u\right) $ only
depends on the variable $u$ and can be constructed with the following set of
expressions 
\begin{eqnarray}
f_{j}\left( u\right) &=&\frac{Z\left( u\right) -Z\left( u_{i}\right) }{\left[
Z\left( u_{j}\right) -Z\left( u_{i}\right) \right] }\ ;\ \ \text{with\ \ }%
i,j=1,2\ \text{ and\ \ }i\neq j  \label{regla de oro} \\
\frac{h_{v}h_{w}}{h_{u}} &\equiv &G\left( u\right) H\left( v,w\right) \ \ \
;\ \ \ \frac{dZ\left( u\right) }{du}\equiv \frac{1}{G\left( u\right) }
\label{regla de oro2}
\end{eqnarray}%
Note that%
\begin{equation}
f_{1}\left( u\right) +f_{2}\left( u\right) =1  \label{f1+f2=1}
\end{equation}%
in the region $V\ $in which this Laplace's equation is satisfied. Such a
property is a particular case of a more general context \cite{capac1}. From
the previous developments, it can be seen that the functions $G\left(
u\right) ,\ H\left( v,w\right) $ and $Z\left( u\right) $ are not unique. We
can redefine such functions as follows%
\begin{equation}
\overline{G}\left( u\right) \equiv \kappa G\left( u\right) \ ,\ \ \bar{H}%
\left( v,w\right) \equiv \frac{1}{\kappa }H\left( v,w\right) \ ,\ \ \ 
\overline{Z}\left( u\right) \equiv \frac{1}{\kappa }Z\left( u\right) +\beta
\label{gauges}
\end{equation}%
where $\kappa $ and $\beta $ are arbitrary constants. Nevertheless, it is
not a problem since these functions are not physical observables. Moreover,
it is easy to check that the BHF's given by Eq. (\ref{regla de oro}) are
invariant under the \textquotedblleft gauge
transformations\textquotedblright\ defined by Eqs. (\ref{gauges}). Finally,
the boundary conditions along with the uniqueness theorems, guarantee that
each solution $f_{j}\left( u\right) $ is unique within the volume $V$.

It worths remarking that the formulas obtained in this section are valid
regardless which surface contains the other. Further, perhaps the most
outstanding property of the BHF's obtained here, is that they depend only on
the geometry. The functions $f_{j}\left( u\right) $ are dimensionless and
given a geometry, they are unique for each $j$.

\subsection{BHF's for a particular case for the scale factors\label{sec:BHF
particular}}

Let us examine the particular case in which%
\begin{equation}
\frac{h_{v}h_{w}}{h_{u}}=1  \label{ansatzh2}
\end{equation}%
in that case from Eq. (\ref{regla de oro2}) the simplest solution for the
functions $G\left( u\right) ,\ H\left( v,w\right) $ and $Z\left( u\right) $
reads%
\begin{equation}
G\left( u\right) =H\left( v,w\right) =1\ \ ;\ \ Z\left( u\right) =u
\label{GHZ simple}
\end{equation}%
Substituting (\ref{GHZ simple}) in (\ref{regla de oro}) the solution for $%
f_{j}\left( u\right) $ simplifies considerably%
\begin{equation}
f_{j}\left( u\right) =\frac{u-u_{i}}{u_{j}-u_{i}}\ ;\ \ \text{with\ \ }%
i,j=1,2\ \text{ and\ \ }i\neq j  \label{HFsimplest}
\end{equation}

The assumption (\ref{ansatzh2}) is fulfilled when we have a system of
coordinates with cylindrical symmetry that is obtained with a conformal
transformation from the cartesian system of coordinates projected onto the $%
XY-$plane. In that case we usually obtain $h_{u}=h_{v}$ and $h_{w}=1$ (of
course the roles of $v$ and $w$ could be interchanged). Note however that a
cylindrical symmetry coming from a conformal transformation is only a
sufficient condition but not necessary. Equation (\ref{ansatzh2}) is the
only condition we require for the validity of Eq. (\ref{HFsimplest}).

\subsection{BHF's for the case of two concentric spheres\label%
{subsec:twospheres1}}

In order to illustrate the method, let us take a geometric configuration
consisting of two concentric spheres of radius $a$ and $b$ with $a>b$.\ We
recall that by definition $u$ is the coordinate such that $u_{i}$ keeps
constant for all points of the surface $S_{i}$ for $i=1,2$. Thus, the
spherical coordinates are the appropriate ones, and $u\equiv r$ is the
coordinate that is kept constant for each of the spherical surfaces. Hence,
for this configuration we have 
\begin{equation}
u=r,\ v=\theta ,\ w=\varphi \ \ \ \text{and\ \ \ }h_{r}=1,\ h_{\theta }=r,\
h_{\varphi }=r\sin \theta  \label{coordin spher}
\end{equation}%
from the first of Eqs. (\ref{regla de oro2}) we can find a solution for $%
G\left( u\right) \equiv G\left( r\right) $ and for $H\left( v,w\right)
\equiv H\left( \theta ,\varphi \right) $%
\begin{equation}
\frac{h_{\theta }h_{\varphi }}{h_{r}}\equiv G\left( r\right) H\left( \theta
,\varphi \right) \ \Rightarrow \ r^{2}\sin \theta =G\left( r\right) H\left(
\theta ,\varphi \right)
\end{equation}%
so a possible solution is\footnote{%
Note that $\bar{G}\left( r\right) =r^{2}$ and $\bar{H}\left( \theta \right)
=\sin \theta $ is also a solution. Both solutions are connected by setting $%
\kappa =-1$ and $\beta =0$\ in\ Eqs. (\ref{gauges}). We prefer solution (\ref%
{Grsolminus}) to obtain a positive solution for $Z\left( u\right) $.}%
\begin{equation}
G\left( r\right) =-r^{2}\ \ ;\ \ H\left( \theta ,\varphi \right) =H\left(
\theta \right) =-\sin \theta  \label{Grsolminus}
\end{equation}%
from the second of Eqs. (\ref{regla de oro2}) we can find a solution for $%
Z\left( u\right) =Z\left( r\right) $ 
\begin{equation}
Z\left( r\right) =-\int \frac{dr}{r^{2}}=\frac{1}{r}  \label{Z(u)spheres}
\end{equation}%
and substituting in Eq. (\ref{regla de oro}) we obtain

\begin{equation}
f_{j}\left( r\right) =\frac{\frac{1}{r}-\frac{1}{r_{i}}}{\left[ \frac{1}{%
r_{j}}-\frac{1}{r_{i}}\right] }\ \ ,\ \ i,j=1,2\text{\ \ \ and\ \ }i\neq j
\label{eq:Esferas}
\end{equation}%
for $r_{j}=r_{2}\equiv a$ and $r_{i}=r_{1}\equiv b$ with $a>b$ Eq. (\ref%
{eq:Esferas}) becomes 
\begin{eqnarray}
f_{2}\left( r\right) &=&\frac{\frac{1}{r}-\frac{1}{r_{1}}}{\left( \frac{1}{%
r_{2}}-\frac{1}{r_{1}}\right) }=\frac{\frac{1}{r}-\frac{1}{b}}{\left( \frac{1%
}{a}-\frac{1}{b}\right) }  \notag \\
f_{2}\left( r\right) &=&\left( \frac{ab}{b-a}\right) \left( \frac{1}{r}-%
\frac{1}{b}\right)  \label{value of fi(r)sphere1}
\end{eqnarray}%
since\ $\nabla ^{2}\left( 1/r\right) =0$ for $r\neq 0$, it is easy to check
that $f_{2}\left( r\right) $ is a solution for Laplace's equation in the
region $b<r<a$ that satisfies the boundary conditions (\ref%
{eq:Condicion_Frontera}) that in our case read 
\begin{equation*}
f_{2}\left( r_{2}\right) =f_{2}\left( a\right) =1\text{\ \ \ and\ \ }%
f_{2}\left( r_{1}\right) =f_{2}\left( b\right) =0
\end{equation*}%
from Eq. (\ref{eq:Esferas}) we can also obtain $f_{1}\left( r\right) $%
\begin{eqnarray}
f_{1}\left( r\right) &=&\frac{\frac{1}{r}-\frac{1}{r_{2}}}{\left( \frac{1}{%
r_{1}}-\frac{1}{r_{2}}\right) }=\frac{\frac{1}{r}-\frac{1}{a}}{\left( \frac{1%
}{b}-\frac{1}{a}\right) }  \notag \\
f_{1}\left( r\right) &=&\left( \frac{ab}{a-b}\right) \left( \frac{1}{r}-%
\frac{1}{a}\right)  \label{value of fi(r)sphere2}
\end{eqnarray}%
which satifies Laplace's equation in the region $b<r<a$ with boundary
conditions $f_{1}\left( a\right) =0$\ and $f_{1}\left( b\right) =1$. Observe
that $f_{1}\left( r\right) +f_{2}\left( r\right) =1$, as it must be.

\subsection{BHF's for two concentric cylinders}

Let us assume two very long concentric cylinders of radii $a$ and $b$ with $%
a>b$. This geometry can be adapted to the polar cylindrical coordinate
system $\left( r,\varphi ,z\right) $. However, we shall use the (equivalent)
conformal polar cylindrical coordinate system $\left( \rho ,\varphi
,z\right) $. Such a system is defined as%
\begin{eqnarray}
x &=&e^{\rho }\cos \varphi \ ;\ \ y=e^{\rho }\sin \varphi \ ;\ \ z=z
\label{conformal cyl} \\
h_{\rho } &=&r\ ;\ \ h_{\varphi }=r\ ;\ \ h_{z}=1  \label{conformal cyl2}
\end{eqnarray}%
the advantage of using the coordinates $\left( \rho ,\varphi ,z\right) $
[instead of the usual cylindrical polar coordinates]\ is that the scale
factors (\ref{conformal cyl2})\ satisfy the condition (\ref{ansatzh2}) from
which it is immediate the form of the BHF's from Eq. (\ref{HFsimplest})%
\begin{equation}
f_{j}\left( \rho \right) =\frac{\rho -\rho _{i}}{\rho _{j}-\rho _{i}}=\frac{%
\ln \left( r/r_{i}\right) }{\ln \left( r_{j}/r_{i}\right) }\ ;\ \ r_{1}=b\
;\ r_{2}=a\ \ \text{with\ \ }i,j=1,2\ \text{ and\ \ }i\neq j
\label{BHF cylinders}
\end{equation}

\section{The case in which the surfaces are \textquotedblleft
equipotential\textquotedblright \label{sec:equipotential case}}

Returning to the general case, it could be realized that owing to the
particular form of the boundary conditions (\ref{eq:Condicion_Frontera}),
the solutions previously generated are remarkably simple and also purely
geometrical as discussed above. We can use the BHF's $f_{j}\left( u\right) $
generated in this way, in order to obtain a more general class of solutions $%
\phi \left( u\right) \ $of Laplace's equation in the same volume $V$, in
which the HF $\phi \left( u\right) \ $acquires constant but arbitrary values 
$\phi _{1}$ and $\phi _{2}$ on the surfaces $S_{1}$ and $S_{2}$
respectively. For such a scenario, the solution $\phi \left( u\right) $ is
given by%
\begin{equation}
\phi \left( u\right) =\phi _{1}f_{1}\left( u\right) +\phi _{2}f_{2}\left(
u\right) =\phi _{2}+f_{1}\left( u\right) ~\left[ \phi _{1}-\phi _{2}\right]
\label{fi(u)gen}
\end{equation}%
where we have used Eq. (\ref{f1+f2=1}) and the functions $f_{j}\left(
u\right) \ $are given by Eqs. (\ref{regla de oro}, \ref{regla de oro2}) in
generalized orthogonal curvilinear coordinates. We can see that Eq. (\ref%
{fi(u)gen}) is the solution to this \textquotedblleft equipotential
problem\textquotedblright\ by observing that if we apply $\nabla ^{2}$ on
both sides of such an equation, we find that $\phi \left( u\right) $ obeys
Laplace's equation. Moreover, by taking a point on the surface $S_{i}$ then $%
u=u_{i}\ $and Eq. (\ref{fi(u)gen}) yields the required boundary conditions%
\begin{equation}
\phi \left( S_{i}\right) =\phi \left( u_{i}\right) =\phi _{i}\ \ \text{with\
\ }i=1,2.  \label{equipot condition}
\end{equation}%
The uniqueness theorems guarantees that such a solution is unique. Note that
in this case the boundary conditions $\phi _{1}$ and $\phi _{2}$ could have
dimensions as well as the general solution $\phi \left( u\right) $. For
instance, in electrostatics $\phi \left( u\right) $ would have dimensions of
potential. This strategy is quite similar to the use of Green's functions to
solve Poisson's equation (or any other linear inhomogeneous equation), in
the sense that Green's functions only depends on the geometry as it is the
case with the BHF's.

As a matter of example, we should keep in mind that the BHF's given by Eqs. (%
\ref{value of fi(r)sphere1}, \ref{value of fi(r)sphere2}) for the concentric
spheres are purely geometrical so far. If we want to solve a physical
problem such as the electrostatic potential in the volume defined by $b<r<a$%
, when the two spherical surfaces are at potentials $\phi _{b}$ and $\phi
_{a}$, such a potential is given by Eq. (\ref{fi(u)gen})%
\begin{equation}
\phi \left( r\right) =\phi _{b}f_{1}\left( r\right) +\phi _{a}f_{2}\left(
r\right) =\phi _{a}+f_{1}\left( r\right) \ \left[ \phi _{b}-\phi _{a}\right]
\ ;\ \ b\leq r\leq a  \label{soluc concentric}
\end{equation}

\section{Formulas of capacitance in orthogonal curvilinear coordinates\label%
{formulas of capacitance}}

\begin{figure}[tbh]
\begin{center}
{\small \includegraphics[width=7.2cm]{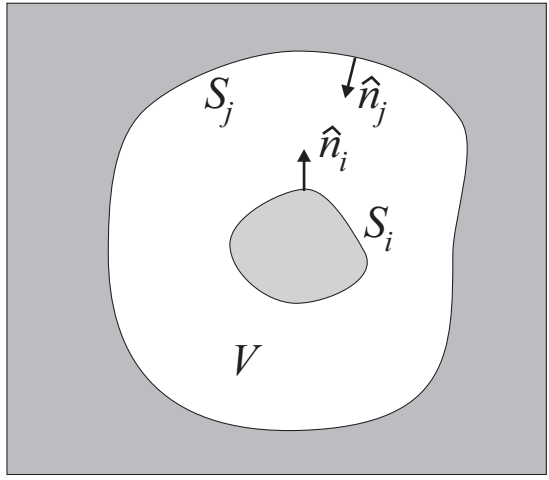} }
\end{center}
\caption{Configuration of two conductors in which the inner conductor with
external surface $S_{i}$ is enclosed within the cavity of an external
conductor. The surface of the cavity is denoted by $S_{j}$,\ and the volume$%
\ V$ is the region (in white) between both surfaces. We could interchange
the role of the labels $i\leftrightarrow j$, and all our developments are
still valid. }
\label{fig:twocond}
\end{figure}

The BHF's described in section \ref{Laplace} have an interesting application
in determining coefficients of capacitance of certain geometrical
configurations. Let us assume that the surfaces $S_{i}$ and $S_{j}$ defined
in section \ref{Laplace} are covered by electrostatic conductors and the
volumen $V$ is left empty. For example, let us assume for a moment that $%
S_{j}$ contains to $S_{i}$. We could have a conductor with a cavity such
that the surface of the cavity is $S_{j}$ and we could have another
conductor inside the cavity whose external surface is $S_{i}$ as displayed
in Fig. \ref{fig:twocond}. Of course, it is possible to interchange the
roles of $S_{i}$ and $S_{j}$ and our developments are still the same.

We can determine the coefficients of capacitance for this geometry of
capacitors by using the BHF's in section \ref{Laplace}. It can be shown that
the coefficients of capacitance for the previous configuration of conductors
are given by \cite{capac1}

\begin{equation}
C_{ij}=-\varepsilon _{0}\oint_{S_{i}}\nabla f_{j}\left( u_{i}\right) \cdot 
\mathbf{n}_{i}dS  \label{eq:CapacitanciaDiazHerrera}
\end{equation}%
where\ $\mathbf{n}_{i}$ is a normal to the surface $S_{i}$ that points
outwards with respect to the $i-th\ $conductor. We have 
\begin{equation}
Q_{i}=\sum_{j}C_{ij}\phi _{j}  \label{potential relation}
\end{equation}%
with $Q_{i},\phi _{i}$ denoting the charge and potential on each conductor.
Expression (\ref{eq:CapacitanciaDiazHerrera}) is valid even if $i=j$.
Nevertheless, we shall consider the case in which $i\neq j$ henceforth.
Since the functions $f_{j}$ take constant values on each surface, then $%
\nabla f_{j}$ evaluated on each surface is perpendicular to such a surface.
Moreover, it can be proved that \cite{capac1}%
\begin{equation}
\nabla f_{j}\left( u_{i},v,w\right) \cdot \mathbf{n}_{i}\geq 0\text{ \ for\ }%
i\neq j  \label{gradtoni}
\end{equation}%
such that $\nabla f_{j}\left( u_{i},v,w\right) $ is parallel (and not
antiparallel) to $\mathbf{n}_{i}$ in each point $u_{i},v,w$ on the surface$\
S_{i}$. From these facts and using Eqs. (\ref{regla de oro}, \ref{regla de
oro2}), as well as the fact that $f_{j}\left( u\right) $ only depends on $u$%
,\ we have%
\begin{eqnarray}
\nabla f_{j}\left( u_{i},v,w\right) \cdot \mathbf{n}_{i} &=&\left\Vert
\nabla f_{j}\left( u_{i},v,w\right) \right\Vert =\left\vert \frac{1}{%
h_{u}\left( u_{i},v,w\right) }\frac{df_{j}\left( u_{i}\right) }{du}%
\right\vert =\left\vert \frac{1}{h_{u}\left( u_{i},v,w\right) ~\left[
Z\left( u_{j}\right) -Z\left( u_{i}\right) \right] }\frac{dZ\left(
u_{i}\right) }{du}\right\vert  \notag \\
\nabla f_{j}\left( u_{i},v,w\right) \cdot \mathbf{n}_{i} &=&\left\vert \frac{%
1}{h_{u}\left( u_{i},v,w\right) ~\left[ Z\left( u_{j}\right) -Z\left(
u_{i}\right) \right] G\left( u_{i}\right) }\right\vert  \label{gradfj}
\end{eqnarray}%
where we have taken into account that all functions involved in integral (%
\ref{eq:CapacitanciaDiazHerrera}) should be evaluated at a given point in $%
S_{i}$ i.e. at $u=u_{i}$. Moreover, since $\mathbf{n}_{i}~dS$ is orthogonal
to the unit vectors $\mathbf{e}_{v}$ and $\mathbf{e}_{w}$\ associated with
the $v,w$ coordinates, we have%
\begin{equation*}
dS=\left\vert h_{v}\left( u_{i},v,w\right) ~h_{w}\left( u_{i},v,w\right)
\right\vert dvdw
\end{equation*}%
where we have taken into account that $dS$ is positive definite. Utilizing
Eqs. (\ref{eq:Ansatz2}) the differential of surface becomes%
\begin{equation}
dS=\left\vert h_{u}\left( u_{i},v,w\right) ~G\left( u_{i}\right) ~H\left(
v,w\right) \right\vert \ dv\ dw  \label{dSinfunctions}
\end{equation}%
substituting (\ref{gradfj}) and (\ref{dSinfunctions})\ in (\ref%
{eq:CapacitanciaDiazHerrera}) the non-diagonal coefficients of capacitance
yield finally 
\begin{equation}
C_{ij}=-\frac{\varepsilon _{0}}{\left\vert Z\left( u_{j}\right) -Z\left(
u_{i}\right) \right\vert }\oint_{S_{i}}\left\vert H\left( v,w\right)
\right\vert dv~dw\ ;\ \ \text{with\ \ }i,j=1,2\ \ \text{and\ \ }i\neq j
\label{eq:Capacitancia_Nueva}
\end{equation}%
we emphasize that despite Eq. (\ref{eq:CapacitanciaDiazHerrera}) is valid
for any values of $i$ and $j$, Eqs. (\ref{gradfj}) and (\ref{gradtoni}) are
only valid for $i\neq j$. In order to obtain the diagonal coefficients we
only have to use the properties \cite{capac1}%
\begin{equation*}
C_{i1}+C_{i2}=0\ ;\ \ C_{ij}=C_{ji}\ \ \Rightarrow \ \ C_{ii}=-C_{ij}\ \ 
\text{with\ \ }i,j=1,2\ \ \text{and\ \ }i\neq j
\end{equation*}%
or more explicitly%
\begin{equation}
C_{11}=C_{22}=-C_{12}=-C_{21}  \label{only one Cij}
\end{equation}%
It worths noting that expression (\ref{eq:Capacitancia_Nueva}) is also
invariant under the \textquotedblleft gauge
transformations\textquotedblright\ given by Eqs. (\ref{gauges}). On the
other hand, Eqs. (\ref{eq:Capacitancia_Nueva}) and (\ref{only one Cij})
leads to \cite{capac1,capac2}%
\begin{equation}
C_{ii}>0\text{ and }C_{ij}<0\text{ \ with\ }i\neq j  \label{CiiCij}
\end{equation}%
Finally, as a proof of consistency, we see from Eq. (\ref{only one Cij})
that the eigenvalues of the 2$\times 2$ matrix of capacitance are $0$ and$\
2C_{11}$. These results along with Eqs. (\ref{CiiCij}) says that the matrix
of capacitance is positive singular as it must be \cite{capac2}. In terms of
voltages Eqs. (\ref{potential relation}, \ref{only one Cij}) becomes%
\begin{equation}
Q_{1}=C_{11}\left( \phi _{1}-\phi _{2}\right) =C_{22}\left( \phi _{1}-\phi
_{2}\right) =-Q_{2}  \label{all charges}
\end{equation}%
Note that if $S_{i}$ refers to the internal conductor, $Q_{i}$ is the total
charge on it. But if $S_{i}$ refers to the cavity of the external conductor
that contains the other conductor, $Q_{i}$ is only the charge accumulated on
the surface of the cavity. In addition, if the internal conductor has a
cavity, the surface of such a cavity does not contribute in the surface
integral to calculate the coefficients of capacitance \cite{Grif}.

\subsection{Capacitance for the particular case of the scale factors}

As it was the case for the solutions of Laplace's equation, the expression
for the coefficients of capacitance is particularly simple when the system
of coordinates obeys relation (\ref{ansatzh2})%
\begin{equation}
\frac{h_{v}h_{w}}{h_{u}}=1  \label{ansatzh}
\end{equation}%
in that case, Eqs. (\ref{GHZ simple}) provide simple solutions for $Z\left(
u\right) $ and $H\left( v,w\right) $. Substituting (\ref{GHZ simple}) in Eq.
(\ref{eq:Capacitancia_Nueva}), the latter becomes 
\begin{equation}
C_{ij}=-\frac{\varepsilon _{0}}{\left\vert u_{j}-u_{i}\right\vert }%
\oint_{S_{i}}dv~dw\ \ ;\ \ i\neq j  \label{eq:Capacitancia_Cilindroij}
\end{equation}%
or for the diagonal elements%
\begin{equation}
C_{ii}=-C_{ij}=\frac{\varepsilon _{0}}{\left\vert u_{j}-u_{i}\right\vert }%
\oint_{S_{i}}dv~dw\ \ ;\ \ \text{with\ \ }i,j=1,2\ \ \text{and\ \ }i\neq j
\label{eq:Capacitancia_Cilindro}
\end{equation}%
We saw that Eq. (\ref{ansatzh}) holds when we have a system of coordinates
with cylindrical symmetry obtained with a conformal transformation from the
cartesian coordinates. Nevertheless, we should recall that a cylindrical
symmetry coming from a conformal transformation is only a sufficient
condition. We only require the condition (\ref{ansatzh}) to guarantee the
validity of Eq. (\ref{eq:Capacitancia_Cilindro}).

Observe that Eq. (\ref{eq:Capacitancia_Cilindro}) has a structure similar to
the capacitance of two parallel plates with \textquotedblleft
area\textquotedblright\ 
\begin{equation}
A_{eff}\equiv \oint_{S_{i}}dv~dw  \label{Aeff}
\end{equation}%
located at a \textquotedblleft distance\textquotedblright\ $\left\vert
u_{j}-u_{i}\right\vert $. Note however that in generalized coordinates, the
integral (\ref{Aeff}) is not necessarily an area. For instance, if the two
coordinates $v$ and $w$ are angular, this quantity is dimensionless. In the
same way, the quantity $\left\vert u_{j}-u_{i}\right\vert $ is not always a
distance.

\subsection{Capacitance of two concentric spheres}

Once again for the sake of illustration, we start calculating the well known
capacitance of two concentric spheres with the values given in subsection %
\ref{subsec:twospheres1}. Thus, the radius of the smaller sphere is$\ b$ and
the radius of the cavity of the bigger sphere is $a$, so that $a>b$.
Applying Eqs. (\ref{Grsolminus}) and (\ref{Z(u)spheres}) in Eq. (\ref%
{eq:Capacitancia_Nueva}) and setting $r_{i}=r_{1}\equiv b$,$\
r_{j}=r_{2}\equiv a\ $we get%
\begin{eqnarray*}
C_{12} &=&-\frac{\varepsilon _{0}}{\left\vert Z\left( r_{2}\right) -Z\left(
r_{1}\right) \right\vert }\oint_{S_{1}}\left\vert H\left( \theta ,\varphi
\right) \right\vert ~d\theta ~d\varphi =-\frac{\varepsilon _{0}}{\left\vert 
\frac{1}{a}-\frac{1}{b}\right\vert }\oint_{S_{1}}\left\vert \sin \theta
\right\vert ~d\theta ~d\varphi \\
C_{12} &=&-\frac{\varepsilon _{0}ab}{\left( a-b\right) }\int_{0}^{\pi }\sin
\theta ~d\theta ~\int_{0}^{2\pi }d\varphi =-\frac{4\pi \varepsilon _{0}ab}{%
\left( a-b\right) }
\end{eqnarray*}%
it is straigthforward that $C_{21}=C_{12}$. Therefore the coefficients of
capacitance yield the expected result%
\begin{equation}
C_{ii}=-C_{ij}=\frac{4\pi \varepsilon _{0}ab}{a-b}\ \ ;\ \ \text{with\ \ }%
i,j=1,2\ \ \text{and\ \ }i\neq j  \label{concentric spheres C}
\end{equation}

\subsection{Capacitance of two concentric cylinders}

As for the case of BHF's, it is more convenient to use the conformal polar
cylindrical coordinates $\left( \rho ,\varphi ,z\right) $ defined by Eqs. (%
\ref{conformal cyl}) instead of the usual polar cylindrical system. Since
for the $\left( \rho ,\varphi ,z\right) $ system the condition (\ref{ansatzh}%
) is fullfilled the capacitance coefficients are given by Eq. (\ref%
{eq:Capacitancia_Cilindro})%
\begin{eqnarray}
C_{ii} &=&\frac{\varepsilon _{0}}{\left\vert \rho _{j}-\rho _{i}\right\vert }%
\oint_{S_{i}}d\varphi ~dz=\frac{\varepsilon _{0}}{\left\vert \rho _{j}-\rho
_{i}\right\vert }\int_{0}^{2\pi }d\varphi ~\int_{-L/2}^{L/2}dz  \notag \\
C_{ii} &=&-C_{ij}=\frac{2\pi \varepsilon _{0}L}{\left\vert \rho _{j}-\rho
_{i}\right\vert }=\frac{2\pi \varepsilon _{0}L}{\left\vert \ln \left(
a/b\right) \right\vert }\ \ ;\ \ i,j=1,2\ \ \text{and\ \ }i\neq j
\label{Capac cylinders}
\end{eqnarray}%
where we have assumed that the cylinders are located in the $z$ interval $%
\left[ -L/2,\ L/2\right] $ with $L>>a$. This is the usual result.

\section{Calculations of BHF's and capacitances in more complex geometries 
\label{capaccalc}}

As we said, the cases of two concentric spheres and two concentric cylinders
were developed for the sake of illustration only. In this section, we shall
calculate BHF's and capacitance coefficients for more complex symmetries in
which our method acquires all its power.

\subsection{\noindent BHF's and capacitance on Prolate spheroidal surfaces}

\begin{figure}[tbh]
\begin{center}
{\small \includegraphics[width=4.2cm]{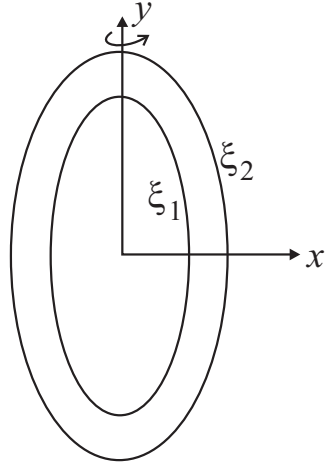} }
\end{center}
\caption{Shell formed between two confocal prolate spheroids.}
\label{fig:prolateshell}
\end{figure}
\noindent

Let us consider a shell formed between two confocal prolate spheroids as
illustrated in Fig.\ \ref{fig:prolateshell}. In order to find the BHF's and
capacitance coefficients, we use the \emph{Prolate spheroidal system of
coordinates}, which is obtained by revolutioning the elliptical
bidimensional system of coordinates $\left( \xi ,\eta \right) \ $with
respect to the major semiaxis of the ellipses. The most usual definition of
the prolate spheroidal system of coordinates reads: 
\begin{eqnarray*}
x &=&a\sinh \xi \sin \eta \cos \varphi \ \ ;\ \ y=a\sinh \xi \sin \eta \sin
\varphi \ \ ;\ \ z=a\cosh \xi \cos \eta \\
\xi &\geq &0,\ \ \eta \in \lbrack 0,\pi ],\ \ \ a=\frac{d}{2}\ \ ;\ \
\varphi \in \lbrack 0,2\pi )
\end{eqnarray*}%
where $\xi ,\eta $ are the bidimensional elliptical coordinates, $\varphi $
is the azimuthal angle and $d$ denotes the distance between the foci. The
scale factors are given by 
\begin{equation}
h_{\xi }=h_{\eta }=a\sqrt{\sinh ^{2}\xi +\sin ^{2}\eta }\ \ ;\ \ h_{\varphi
}=a\sinh \xi \sin \eta  \label{prolate coordh}
\end{equation}%
By means of the trigonometric and hyperbolic identities 
\begin{eqnarray}
\frac{z^{2}}{a^{2}\cosh ^{2}\xi }+\frac{x^{2}+y^{2}}{a^{2}\sinh ^{2}\xi }
&=&\cos ^{2}\eta +\sin ^{2}\eta =1  \label{hiper id pro1} \\
\frac{z^{2}}{a^{2}\cos ^{2}\eta }-\frac{x^{2}+y^{2}}{a^{2}\sin ^{2}\eta }
&=&\cosh ^{2}\xi -\sinh ^{2}\xi =1  \label{hiper id pro2}
\end{eqnarray}%
it can be seen that the constant values of $\xi $ correspond to prolate
spheroids, while surfaces of constant $\eta $ are hyperboloids of
revolution. Consequently, since we are interested in the configuration in
which the surfaces are confocal prolate spheroids, the coordinate that is
kept constant is $\xi $. Then we shall assign $\left( u,v,w\right)
\rightarrow \left( \xi ,\eta ,\varphi \right) $. Combining Eqs. (\ref%
{prolate coordh}) and (\ref{regla de oro2}) we have%
\begin{equation*}
\frac{h_{\eta }h_{\varphi }}{h_{\xi }}=h_{\varphi }=a\sinh \xi \sin \eta
\equiv G\left( \xi \right) H\left( \eta ,\varphi \right)
\end{equation*}%
Such that a set of solutions for $G\left( u\right) $ and $H\left( v,w\right) 
$ is given by%
\begin{equation}
G\left( \xi \right) =a\sinh \xi \ \ ;\ \ H\left( \eta ,\varphi \right) =\sin
\eta  \label{GH prolate}
\end{equation}%
and from Eqs. (\ref{regla de oro2}, \ref{GH prolate}), the simplest solution
for $Z\left( u\right) $ yields%
\begin{eqnarray}
Z\left( \xi \right) &=&\int \frac{d\xi }{G\left( \xi \right) }=\frac{1}{a}%
\int \frac{1}{\sinh \xi }\ d\xi  \notag \\
Z\left( \xi \right) &=&\frac{1}{a}\ln \left( \tanh \frac{\xi }{2}\right)
\label{Z prolate}
\end{eqnarray}%
substituting (\ref{Z prolate}) in (\ref{regla de oro}) we obtain%
\begin{eqnarray}
f_{j}\left( \xi \right) &=&\frac{Z\left( \xi \right) -Z\left( \xi
_{i}\right) }{\left[ Z\left( \xi _{j}\right) -Z\left( \xi _{i}\right) \right]
}  \notag \\
f_{j}\left( \xi \right) &=&\frac{\ln \left( \tanh \frac{\xi }{2}\right) -\ln
\left( \tanh \frac{\xi _{i}}{2}\right) }{\ln \left( \tanh \frac{\xi _{j}}{2}%
\right) -\ln \left( \tanh \frac{\xi _{i}}{2}\right) }=\frac{\ln \left( \frac{%
\tanh \frac{\xi }{2}}{\tanh \frac{\xi _{i}}{2}}\right) }{\ln \left( \frac{%
\tanh \frac{\xi _{j}}{2}}{\tanh \frac{\xi _{i}}{2}}\right) }\ \ ;\ \ i,j=1,2%
\text{\ \ and\ \ }i\neq j  \label{f1prolate}
\end{eqnarray}%
where $\xi _{i}$ and $\xi _{j}$ are the constant values of that coordinate
on the surfaces\ $S_{i}$ and $S_{j}$. Now for the capacitance we subtitute (%
\ref{GH prolate}) and (\ref{Z prolate}) in (\ref{eq:Capacitancia_Nueva}) to
get%
\begin{eqnarray*}
C_{12} &=&-\frac{\varepsilon _{0}}{\left\vert Z\left( \xi _{2}\right)
-Z\left( \xi _{1}\right) \right\vert }\oint_{S_{1}}\left\vert H\left( \eta
,\varphi \right) \right\vert ~d\eta ~d\varphi \\
C_{12} &=&-\frac{\varepsilon _{0}}{\left\vert \frac{1}{a}\ln \left( \tanh 
\frac{\xi _{2}}{2}\right) -\frac{1}{a}\ln \left( \tanh \frac{\xi _{1}}{2}%
\right) \right\vert }\int_{0}^{2\pi }~d\varphi \int_{0}^{\pi }\sin \eta
~d\eta
\end{eqnarray*}%
the explicit calculation of $C_{21}$ gives $C_{21}=C_{12}$ as it must be. So
we finally obtain%
\begin{equation}
C_{ii}=-C_{ij}=\frac{4\pi a\varepsilon _{0}}{\left\vert \ln \left( \frac{%
\tanh \frac{\xi _{j}}{2}}{\tanh \frac{\xi _{i}}{2}}\right) \right\vert }\ \
;\ \ i,j=1,2\text{\ \ and\ \ }i\neq j  \label{capacprolate}
\end{equation}

The reader can contrast the simplicity and brevity of the calculation in
this subsection, with respect to the methods developed in the literature
(see for instance Ref. \cite{geomcapac1}). Once the general formulas (\ref%
{regla de oro}, \ref{regla de oro2}) and (\ref{eq:Capacitancia_Nueva}) are
obtained, the application to this specific geometry is just a bit more than
a direct replacement. This is also the case with other symmetries as we
shall see.

Finally, it is desirable to obtain the capacitance in terms of geometrical
factors. Defining $a_{i}$ as the major semiaxes of the prolate spheroid
associated with $\xi _{i}$, the capacitance (\ref{capacprolate}) becomes
(see appendix \ref{ap:coordinate geom})%
\begin{equation}
C_{ii}=-C_{ij}=\frac{4\pi a\varepsilon _{0}}{\left\vert \ln \left( \frac{%
\left( a+a_{i}\right) \sqrt{a_{j}^{2}-a^{2}}}{\left( a+a_{j}\right) \sqrt{%
a_{i}^{2}-a^{2}}}\right) \right\vert }\ \ ;\ \ i,j=1,2\text{\ \ and\ \ }%
i\neq j  \label{capacprolate2}
\end{equation}%
it is also shown in appendix \ref{ap:coordinate geom} that the capacitance
of two concentric spheres is obtained from the limit where $a$ tends to zero.

\subsection{BHF's and capacitance on oblate spheroidal surfaces}

\begin{figure}[tbh]
\begin{center}
{\small \includegraphics[width=4.2cm]{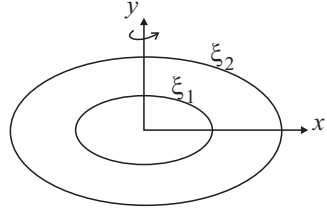} }
\end{center}
\caption{A shell formed between two confocal oblate spheroidal surfaces.}
\label{fig:oblateshell}
\end{figure}

Let us consider a shell formed between two confocal oblate spheroids as
displayed in Fig. \ref{fig:oblateshell}. Then, we should work with the \emph{%
oblate spheroidal coordinate system}\noindent\ which is obtained by
revolutioning the bidimensional elliptical coordinate system $\left( \xi
,\eta \right) \ $around the minor semiaxis of the ellipses. The
transformation of coordinates and scale factors yield 
\begin{eqnarray}
x &=&a\cosh \xi \cos \eta \cos \varphi \ \ ;\ \ y=a\cosh \xi \cos \eta \sin
\varphi \ \ ;\ \ z=a\sinh \xi \sin \eta  \label{coordinates oblate} \\
h_{\xi } &=&h_{\eta }=a\sqrt{\sinh ^{2}\xi +\sin ^{2}\eta }\ \ ;\ \
h_{\varphi }=a\cosh \xi \cos \eta  \label{scale factors oblate} \\
\xi &\geq &0,\ \ \eta \in \lbrack -\frac{\pi }{2},\frac{\pi }{2}],\ \ \ a=%
\frac{d}{2}\ \ ;\ \ \varphi \in \lbrack 0,2\pi )  \label{intervals oblate}
\end{eqnarray}%
The trigonometric and hyperbolic identities 
\begin{eqnarray}
\frac{z^{2}}{a^{2}\sinh ^{2}\xi }+\frac{x^{2}+y^{2}}{a^{2}\cosh ^{2}\xi }
&=&\cos ^{2}\eta +\sin ^{2}\eta =1  \label{ec elipsoid obl2} \\
\frac{x^{2}+y^{2}}{a^{2}\cos ^{2}\eta }-\frac{z^{2}}{a^{2}\sin ^{2}\eta }
&=&\cosh ^{2}\xi -\sinh ^{2}\xi =1  \label{ec elipsoid obl2b}
\end{eqnarray}%
show that the surfaces in which $\xi \ $acquires constant values correspond
to oblate spheroids, while the surfaces with $\eta $ constant correspond to
hyperboloids of revolution. Once again, $\xi $ is the variable that is kept
constant in the surfaces we are working with. Equations (\ref{regla de oro2}%
) and (\ref{scale factors oblate}) are combined to give%
\begin{eqnarray}
\frac{h_{\eta }h_{\varphi }}{h_{\xi }} &=&h_{\varphi }=a\cosh \xi \cos \eta
\equiv G\left( \xi \right) H\left( \eta ,\varphi \right)  \notag \\
G\left( \xi \right) &\equiv &a\cosh \xi \ \ ;\ \ H\left( \eta ,\varphi
\right) \equiv \cos \eta  \label{GH oblate} \\
Z\left( \xi \right) &=&\frac{1}{a}\int \frac{1}{\cosh \xi }d\xi =\frac{1}{a}%
\sin ^{-1}\left( \tanh \xi \right)  \label{Z oblate}
\end{eqnarray}%
Using Eqs. (\ref{GH oblate}, \ref{Z oblate}) in (\ref{regla de oro}) we
obtain%
\begin{equation}
f_{j}\left( \xi \right) =\frac{Z\left( \xi \right) -Z\left( \xi _{i}\right) 
}{\left[ Z\left( \xi _{j}\right) -Z\left( \xi _{i}\right) \right] }=\frac{%
\sin ^{-1}\left( \tanh \xi \right) -\sin ^{-1}\left( \tanh \xi _{i}\right) }{%
\sin ^{-1}\left( \tanh \xi _{j}\right) -\sin ^{-1}\left( \tanh \xi
_{i}\right) }\ \ ;\ \ i,j=1,2\text{\ \ and\ \ }i\neq j  \label{BHF oblate}
\end{equation}%
and the capacitance coefficients are obtained from (\ref{GH oblate}, \ref{Z
oblate}) and (\ref{eq:Capacitancia_Nueva})%
\begin{eqnarray}
C_{ij} &=&\frac{-\varepsilon _{0}}{\left\vert Z\left( \xi _{j}\right)
-Z\left( \xi _{i}\right) \right\vert }\oint_{S_{1}}\left\vert H\left( \eta
,\varphi \right) \right\vert ~d\eta ~d\varphi =\frac{-\varepsilon _{0}}{%
\left\vert \frac{1}{a}\sin ^{-1}\left( \tanh \xi _{j}\right) -\frac{1}{a}%
\sin ^{-1}\left( \tanh \xi _{i}\right) \right\vert }\int_{0}^{2\pi }d\varphi
\int_{-\pi /2}^{\pi /2}\cos \eta ~d\eta  \notag \\
C_{ii} &=&-C_{ij}=\frac{4\pi a\varepsilon _{0}}{\left\vert \sin ^{-1}\left(
\tanh \xi _{j}\right) -\sin ^{-1}\left( \tanh \xi _{i}\right) \right\vert }\
\ ;\ \ i,j=1,2\text{\ \ and\ \ }i\neq j  \label{oblategeo}
\end{eqnarray}%
for instance, if we compare with the procedure in Ref. \cite{geomcapac3b},
the advantages of our method are apparent. In terms of geometrical factors
the coefficients $C_{ij}$ read (see appendix \ref{ap:coordinate geom})%
\begin{equation}
C_{ii}=-C_{ij}=\frac{4\pi a\varepsilon _{0}}{\left\vert \cos ^{-1}\epsilon
_{j}-\cos ^{-1}\epsilon _{i}\right\vert }\ \ ;\ \ i,j=1,2\text{\ \ and\ \ }%
i\neq j  \label{oblategeom}
\end{equation}%
where $\epsilon _{j}$,$\ \epsilon _{i}$ are the eccentricities of the
ellipses associated with the surfaces $S_{j}$ and $S_{i}$ respectively.

\subsection{BHF's and capacitance of two confocal elliptical cylinders}

\begin{figure}[tbh]
\begin{center}
{\small \includegraphics[width=4.2cm]{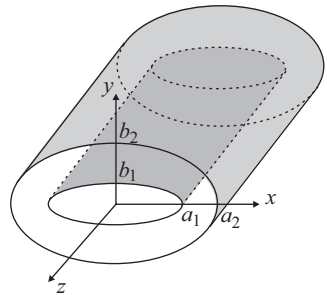} }
\end{center}
\caption{A shell formed between two confocal elliptical cylinders. The
values $a_{1},~a_{2}$ correspond to the major semiaxes, while $b_{1},~b_{2}$
denote the minor semiaxes.}
\label{fig:confocal cylinders}
\end{figure}

This is an example in which the particular condition (\ref{ansatzh2}) is
fulfilled. Let us consider a shell formed by two confocal elliptical
cylinders as shown in Fig. \ref{fig:confocal cylinders}. The values $b_{1}$, 
$a_{1}$, correspond to the minor semiaxis and major semiaxis of the internal
cylinder, while $b_{2}$, $a_{2}$ are the semiaxes associated with the
external cylinder. We assume that along the $Z-$axis the cylinders are
located in the interval $\left[ -L/2,~L/2\right] $ with $L>>a_{2}$. We shall
use the\emph{\ cylindrical elliptical system of coordinates} which is
obtained by projecting the elliptical bidimensional system on the $Z-$axis.
The transformation of coordinates and scale factors yield 
\begin{eqnarray}
x &=&a\cosh \xi \cos \eta \ ;\ y=a\sinh \xi \sin \eta \ \ ;\ z=z
\label{elipcylinder1} \\
h_{\xi } &=&h_{\eta }=a\sqrt{\sinh ^{2}\xi +\sin ^{2}\eta }\ \ ;\ \ h_{z}=1
\label{elipcylinder2} \\
\xi &\geq &0\ ;\ \ 0\leq \eta <2\pi \ ;\ \ -\infty <z<\infty \ ;\ \ a=\frac{d%
}{2}
\end{eqnarray}%
The trigonometrical and hyperbolic identities 
\begin{eqnarray}
\frac{x^{2}}{a^{2}\cosh ^{2}\xi }+\frac{y^{2}}{a^{2}\sinh ^{2}\xi } &=&\cos
^{2}\eta +\sin ^{2}\eta =1  \label{elipcyl3} \\
\frac{x^{2}}{a^{2}\cos ^{2}\eta }-\frac{y^{2}}{a^{2}\sin ^{2}\eta } &=&\cosh
^{2}\xi -\sinh ^{2}\xi =1  \label{elipcyl4}
\end{eqnarray}%
say that surfaces with constant $\xi $ are confocal elliptical cylinders and
surfaces with $\eta $ constant are confocal hyperbolic cylinders. Thus $\xi $
is the coordinate that takes constant values in the surfaces we are
considering. Equation (\ref{elipcylinder2}) shows that in this case, the
condition (\ref{ansatzh2}) is satisfied. Consequently, the BHF's are given
by Eq. (\ref{HFsimplest}) that in this case becomes

\noindent 
\begin{equation}
f_{j}\left( \xi \right) =\frac{\xi -\xi _{i}}{\xi _{j}-\xi _{i}}\ \ ;\ \
i,j=1,2\text{\ \ and\ \ }i\neq j  \label{HFellipsecylinder}
\end{equation}%
and the coefficients of capacitance are given by Eq. (\ref%
{eq:Capacitancia_Cilindro}) that in our case gives%
\begin{eqnarray}
C_{11} &=&-C_{12}=\frac{\varepsilon _{0}}{\left\vert \xi _{2}-\xi
_{1}\right\vert }\oint_{S_{1}}d\eta \ dz=\frac{\varepsilon _{0}}{\left\vert
\xi _{2}-\xi _{1}\right\vert }\int_{0}^{2\pi }d\eta \int_{-L/2}^{L/2}dz 
\notag \\
C_{ii} &=&-C_{ij}=\frac{2\pi \varepsilon _{0}L}{\left\vert \xi _{i}-\xi
_{j}\right\vert }\ \ ;\ \ i,j=1,2\text{\ \ and\ \ }i\neq j
\label{cijcylindellip1}
\end{eqnarray}%
Note that the integral has the same value if we integrate over $S_{2}$, it
guarantees that $C_{12}=C_{21}$. This example shows that under the condition
(\ref{ansatzh2}) the results are much simpler. In terms of geometrical
factors Eq. (\ref{cijcylindellip1}) becomes (see appendix \ref{ap:coordinate
geom})%
\begin{equation}
C_{ii}=-C_{ij}=\frac{2\pi \varepsilon _{0}L}{\left\vert \ln \left[ \frac{%
a_{i}+\sqrt{a_{i}^{2}-a^{2}}}{a_{j}+\sqrt{a_{j}^{2}-a^{2}}}\right]
\right\vert }\ \ ;\ \ i,j=1,2\text{\ \ and\ \ }i\neq j
\label{cijcylindellip2}
\end{equation}%
the limit $a\rightarrow 0$, reduces to the case of two concentric cylinders.

\subsection{BHF's and capacitance of two eccentric cylinders}

\begin{figure}[tbh]
\begin{center}
{\small \includegraphics[width=6cm]{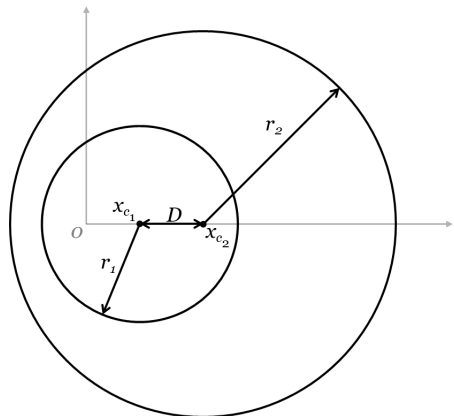} }
\end{center}
\caption{Two eccentric cylinders $C_{1}$ and $C_{2}\ $with radii $r_{1}$ and 
$r_{2}$. The cylinder $C_{1}$ is contained in the cylinder $C_{2}$, and
their centers are separated by a distance $D$.}
\label{fig:bipolar}
\end{figure}
\begin{figure}[tbh]
\begin{center}
{\small \includegraphics[width=9cm]{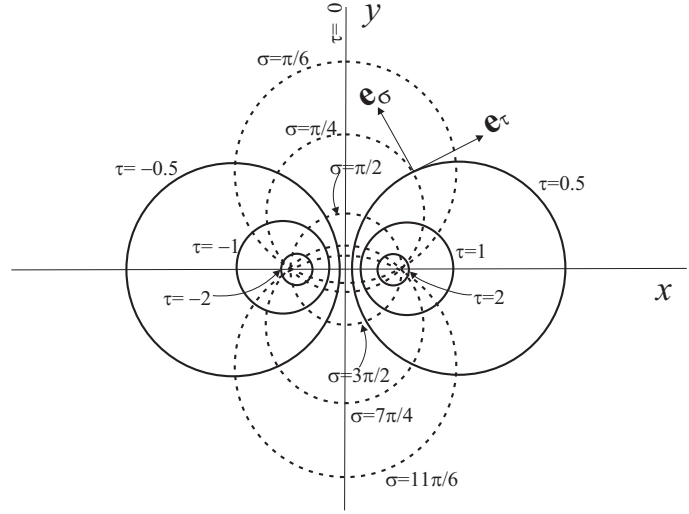} }
\end{center}
\caption{Bipolar coordinate system (in two dimensions) with foci $%
F_{1}=\left( 0,-a\right) $ and $F_{2}=\left( 0,a\right) $. It is shown that
curves associated with positive (negative)\ constant values of $\protect\tau 
$ are circles that contain the focus $F_{2}\ $($F_{1}$). When $\left\vert 
\protect\tau _{1}\right\vert >\left\vert \protect\tau _{2}\right\vert $ the
circle $C_{2}$ contains the circle $C_{1}$.}
\label{fig:bipolar1}
\end{figure}

Let us consider a shell like the one in Fig. \ref{fig:bipolar}. The external
cylinder has a radius $r_{2}$, and the internal one a radius $r_{1}$. The
centers are located at the points $\left( x_{c_{1}},0\right) $ and $\left(
x_{c_{2}},0\right) $, and are separated by a distance $D$. The cylindrical
bipolar coordinate system is the appropriate for this geometry. The
(bidimensional) bipolar coordinate system posseses two foci $F_{1}$ and $%
F_{2}$, usually located at $(-a,0)$ and $(a,0)$ respectively. The
transformations from cartesian coordinates and scale factors are given by 
\begin{eqnarray*}
x &=&a\frac{\sinh \tau }{\cosh \tau -\cos \sigma }\ \ ;\ \ y=a\frac{\sin
\sigma }{\cosh \tau -\cos \sigma }\ \ ;\ \ z=z \\
h_{\sigma } &=&h_{\tau }=\frac{a}{\cosh \tau -\cos \sigma }\ ;\ h_{z}=1 \\
0 &\leq &\sigma <2\pi \ \ ;\ \ -\infty <\tau <+\infty \ \ ;\ \ -\infty
<z<+\infty
\end{eqnarray*}%
The distance from each focus to an arbitrary point $\left( x,y\right) \ $in
the plane yields%
\begin{equation*}
d_{1}^{2}=(x+a)^{2}+y^{2}\ \ ;\ d_{2}^{2}=(x-a)^{2}+y^{2}
\end{equation*}%
Thus, $\sigma $ represents the angle between the lines that join each focus
with a given point $\left( x,y\right) $ in the plane, while $\tau $ is given
by%
\begin{equation*}
\tau =\ln \left( \frac{d_{1}}{d_{2}}\right)
\end{equation*}%
the curves (in the plane) with $\sigma $ constant correspond to
non-concentric circles that intersect in the two foci (see Fig. \ref%
{fig:bipolar1}). The curves with $\tau $ constant correspond to
non-intersecting circles and with different radii (see Fig. \ref%
{fig:bipolar1}). Two curves $C_{1}$ and $C_{2}\ $with constant positive
values $\tau _{1}>\tau _{2}$ correspond to two non-concentric circles such
that both circles contain the focus $F_{2}$ and the circle $C_{2}$ contains
the circle $C_{1}$. Adding the $z$ coordinate, we obtain the desire
configuration of two non-concentric cylinders\footnote{%
We can choose negative values of $\tau $ such that the circles contain the
focus $F_{1}$.}. Therefore, we assign $\left( u,v,w\right) \rightarrow
\left( \tau ,\sigma ,z\right) $. Note that this system of coordinates
satisfy condition (\ref{ansatzh2}) so that the solutions for the
coefficients of capacitance and BHF's are given by Eqs. (\ref{HFsimplest})
and (\ref{eq:Capacitancia_Cilindro})%
\begin{eqnarray}
f_{j}\left( \tau \right) &=&\frac{\tau -\tau _{i}}{\tau _{j}-\tau _{i}}\ \
;\ \ i,j=1,2\text{ and }i\neq j  \label{CIIeccentric2} \\
C_{ii} &=&\frac{\varepsilon _{0}}{\left\vert \tau _{j}-\tau _{i}\right\vert }%
\oint_{S_{i}}d\sigma ~dz=\frac{\varepsilon _{0}}{\left\vert \tau _{j}-\tau
_{i}\right\vert }\int_{0}^{2\pi }d\sigma ~\int_{-L/2}^{L/2}dz  \notag \\
C_{ii} &=&-C_{ij}=\frac{2\pi \varepsilon _{0}L}{\left\vert \tau _{j}-\tau
_{i}\right\vert }\ \ ;\ \ i,j=1,2\text{ and }i\neq j  \label{CIIeccentric1}
\end{eqnarray}

Substituting Eq. (\ref{taosgeom}) of appendix \ref{ap:coordinate geom} in
Eq. (\ref{CIIeccentric1}), we obtain the coefficients of capacitance in
terms of geometrical factors%
\begin{equation}
C_{ii}=-C_{ij}=\frac{2\pi \varepsilon _{0}L}{\cosh ^{-1}\left( \frac{%
r_{1}^{2}+r_{2}^{2}-D^{2}}{2r_{1}r_{2}}\right) }  \label{CIIeccentric3}
\end{equation}%
when $D\rightarrow 0$ we recover the expression for two concentric cylinders.

\section{Conclusions\label{conclusions}}

We solve Laplace's equation for some geometrical configurations in which
there exist a system of coordinates $\left( u,v,w\right) \ $that can be
adjusted to the geometry. Specifically, we assume that the boundary consists
of two closed surfaces $S_{1}$ and $S_{2}$ so that one of the surfaces is
included into the other, and in which the coordinate $u\ $takes constant
values $u_{1}$ and $u_{2}$ on each surface $S_{1}$ and $S_{2}$. We solve
Laplace's equation within the volume $V$ between both surfaces. Besides, we
take specific boundary conditions in which the solutions take values that
are either zero or unity on the surfaces. The solutions of Laplace's
equation with such specific boundary conditions are called Basic Harmonic
Functions (BHF's). It worths remarking that the BHF's only depend on the
geometry, and their solutions via separation of variables are particularly
simple. The expressions for the BHF's are given in terms of generalized
curvilinear coordinates in such a way that we can sweep a variety of systems
of coordinates to obtain the BHF's associated with the given symmetry.

Further, we exploit the purely geometrical BHF's in two ways. On one hand,
we can determine a more general class of Harmonic Functions (HF's i.e.
solutions of Laplace's equation) for the same geometry of the BHF's, but in
which the functions take arbitrary constant values on the boundaries. On the
other hand, we calculate the coefficients of capacitance when the surfaces
that form the volume are covered with electrostatic conductors. Expressions
for either HF's and for coefficients of capacitance are given again in terms
of generalized curvilinear coordinates, such that we can obtain these
quantities for multiple symmetries by simple replacements. Several examples
with different symmetries are provided.

In addition, appendices \ref{sec:field density} and \ref{sec:embedding} show
natural extensions coming from our present formulation. In appendix \ref%
{sec:field density}, expressions for the electric field between two
conductors and the charge densities on each of their surfaces are written in
generalized curvilinear coordinates, while in appendix \ref{sec:embedding},
formulas for the capacitance coefficients are extended to the case of
conductors in sucessive embedding.

It worths emphasizing that after obtaining the general formulas for BHF's,
capacitance coefficients, electric fields and surface charge densities in
terms of generalized orthogonal curvilinear coordinates, calculations of
these observables for each symmetry are straightforward. After checking Eqs.
(\ref{regla de oro}, \ref{regla de oro2}, \ref{eq:Capacitancia_Nueva}) and
Eqs. (\ref{electric curvilinear}, \ref{regla de oro surface}) we can see
that all those observables are obtained by simple combinations of the
functions $h_{u},G\left( u\right) ,\ H\left( v,w\right) $ and $Z\left(
u\right) $. The first three are basically input parameters, and the latter
is obtained from $1/G\left( u\right) $ by a direct integration. Comparison
with the calculations of the same quantities with other methods given in the
literature, shows the simplicity and brevity of our method as it was
emphasized in several of our examples.

Note that despite with the traditional strategies Laplace's Equation can be
solved in many cases via separation of variables, in most of such cases a
series expansion in certain special functions is usually required in order
to adjust the boundary conditions. By contrast, the BHF's obtained via the
formulas (\ref{regla de oro}, \ref{regla de oro2}) lead in many problems to
closed solutions. Therefore, solutions of Laplace's equation with
equipotential surfaces can often be written (with our method) in terms of
closed well-defined functions. Though solutions in series are exact in
principle, when we handle them in practice we should truncate the series
somewhere, leading to a considerable expansion of numerical errors,
especially if the series converges slowly (sometimes problems of convergence
also arise).

On the other hand, it is well known that solutions of Laplace's equation can
be treated with the geometrical formalism of Green's functions.
Nevertheless, it could be noticed that finding BHF's (which are also purely
geometrical)\ is in general much easier than the calculation of the Green's
function on the same geometry. Finally, it is possible to extrapolate the
method developed here to other linear homogeneous differential equations. In
the same way that we can associate a Green's function with each linear
differential operator, we can define BHF's for each linear differential
operator.

\section*{Acknowledgments}

We acknowledge to división de investigación de Bogotá (DIB) of Universidad
Nacional de Colombia for its financial support.

{\small \appendix}

\section{Electric field and surface density in a configuration of two
conductors\label{sec:field density}}

Once again we use $i,j=1,2$ and $i\neq j$ throughout this section. If $\phi
_{i}$ and $\phi _{j}$ are the (constant) potentials on conductors $i$ and $j$
respectively, the potential in the volume $V$ between the surfaces $S_{i}$
and $S_{j}$ is given by Eq. (\ref{fi(u)gen})%
\begin{equation}
\phi =\phi _{i}f_{i}+\phi _{j}f_{j}  \label{fi(u)gen2}
\end{equation}%
on the other hand, from the property $f_{i}+f_{j}=1$ we obtain%
\begin{equation}
\nabla f_{i}=-\nabla f_{j}  \label{gradfifj}
\end{equation}%
combining Eqs. (\ref{fi(u)gen2}, \ref{gradfifj}) and the fact that the BHF's
only depend on $u$, we can find the electric field within the volume $V\ $in
terms of curvilinear coordinates%
\begin{equation*}
\mathbf{E}=-\nabla \phi =-\phi _{i}\nabla f_{i}-\phi _{j}\nabla f_{j}=\left(
\phi _{i}-\phi _{j}\right) \nabla f_{j}=\frac{\phi _{i}-\phi _{j}}{%
h_{u}\left( u,v,w\right) }\frac{\partial f_{j}\left( u\right) }{\partial u}%
\mathbf{e}_{u}\left( u,v,w\right)
\end{equation*}%
where $\mathbf{e}_{u}\left( u,v,w\right) $ is the local unitary vector
associated with the $u\ $coordinate. Using Eqs. (\ref{regla de oro}, \ref%
{regla de oro2}) we have%
\begin{eqnarray}
\mathbf{E}\left( u,v,w\right) &=&\frac{\phi _{i}-\phi _{j}}{h_{u}\left(
u,v,w\right) ~\left[ Z\left( u_{j}\right) -Z\left( u_{i}\right) \right] }%
\frac{dZ\left( u\right) }{du}\mathbf{e}_{u}\left( u,v,w\right)  \notag \\
\mathbf{E}\left( u,v,w\right) &=&\frac{\phi _{i}-\phi _{j}}{h_{u}\left(
u,v,w\right) ~\left[ Z\left( u_{j}\right) -Z\left( u_{i}\right) \right]
~G\left( u\right) }\mathbf{e}_{u}\left( u,v,w\right) \ \ ;\ \ i,j=1,2\ \ 
\text{and }i\neq j  \label{electric curvilinear}
\end{eqnarray}

On the other hand, the surface charge density induced on the surface $S_{i}$
of a perfect conductor is given by \cite{Grif}%
\begin{equation*}
\sigma \left( S_{i}\right) =\sigma \left( u_{i},v,w\right) =-\varepsilon
_{0}\nabla \phi \left( u_{i},v,w\right) \cdot \mathbf{n}_{i}=\varepsilon
_{0}\left( \phi _{i}-\phi _{j}\right) \nabla f_{j}\left( u_{i},v,w\right)
\cdot \mathbf{n}_{i}
\end{equation*}%
where we have used Eqs. (\ref{fi(u)gen2}) and (\ref{gradfifj}). Finally, by
using Eq. (\ref{gradfj}) we have%
\begin{equation}
\sigma \left( u_{i},v,w\right) =\left\vert \frac{\varepsilon _{0}}{%
h_{u}\left( u_{i},v,w\right) ~\left[ Z\left( u_{j}\right) -Z\left(
u_{i}\right) \right] G\left( u_{i}\right) }\right\vert \left( \phi _{i}-\phi
_{j}\right) \ \ ;\ \ i,j=1,2\text{\ and\ }i\neq j
\label{regla de oro surface}
\end{equation}

As a proof of consistency, let us obtain the total charge on the surface $%
S_{i}$. By using Eqs. (\ref{regla de oro surface}, \ref{dSinfunctions}) and (%
\ref{eq:Capacitancia_Nueva}, \ref{only one Cij}) we obtain%
\begin{eqnarray}
Q_{i} &=&\oint_{S_{i}}\sigma \left( u_{i},v,w\right) ~dS=\varepsilon
_{0}\left( \phi _{i}-\phi _{j}\right) \oint_{S_{i}}\left\vert \frac{dS}{%
h_{u}\left( u_{i},v,w\right) ~\left[ Z\left( u_{j}\right) -Z\left(
u_{i}\right) \right] G\left( u_{i}\right) }\right\vert  \notag \\
&=&\varepsilon _{0}\left( \phi _{i}-\phi _{j}\right) \oint_{S_{i}}\left\vert 
\frac{h_{u}\left( u_{i},v,w\right) ~G\left( u_{i}\right) ~H\left( v,w\right) 
}{h_{u}\left( u_{i},v,w\right) ~\left[ Z\left( u_{j}\right) -Z\left(
u_{i}\right) \right] G\left( u_{i}\right) }\right\vert \ dv\ dw  \notag \\
&=&\left( \phi _{i}-\phi _{j}\right) \frac{\varepsilon _{0}}{\left\vert
Z\left( u_{j}\right) -Z\left( u_{i}\right) \right\vert }\oint_{S_{i}}\left%
\vert H\left( v,w\right) \right\vert \ dv\ dw=-C_{ij}\left( \phi _{i}-\phi
_{j}\right)  \notag \\
Q_{i} &=&C_{ii}\left( \phi _{i}-\phi _{j}\right)  \label{Q=CVnow}
\end{eqnarray}%
which reproduces the correct expression (\ref{all charges}) for the total
charge on $S_{i}$. If we compare expressions (\ref{regla de oro surface})
and (\ref{Q=CVnow}), the term in the absolute value of Eq. (\ref{regla de
oro surface}) behaves like an \textquotedblleft effective capacitance per
unit area\textquotedblright . Effectively, such a term is purely
geometrical, as is the term $C_{ii}$.

\subsection{Electric fields and charge densities for two concentric spheres}

For the two concentric spheres of Sec. \ref{subsec:twospheres1}, we
substitute Eqs. (\ref{coordin spher}, \ref{Grsolminus}, \ref{Z(u)spheres})
in Eq. (\ref{electric curvilinear}) to obtain%
\begin{eqnarray}
\mathbf{E}\left( \mathbf{r}\right) &=&\frac{\left( \phi _{i}-\phi
_{j}\right) \mathbf{e}_{r}}{h_{r}~\left[ Z\left( r_{j}\right) -Z\left(
r_{i}\right) \right] ~G\left( r\right) }=\frac{\left( \phi _{1}-\phi
_{2}\right) \mathbf{e}_{r}}{\left[ \frac{1}{r_{2}}-\frac{1}{r_{1}}\right]
~\left( -r^{2}\right) }=\frac{\left( \phi _{b}-\phi _{a}\right) \mathbf{e}%
_{r}}{\left[ \frac{1}{a}-\frac{1}{b}\right] ~\left( -r^{2}\right) }  \notag
\\
\mathbf{E}\left( \mathbf{r}\right) &=&\frac{ab\left( \phi _{b}-\phi
_{a}\right) }{\left( a-b\right) ~r^{2}}\mathbf{e}_{r}
\label{concentric spheres E}
\end{eqnarray}%
and substituting (\ref{coordin spher}, \ref{Grsolminus}, \ref{Z(u)spheres})
in Eq. (\ref{regla de oro surface}) we find%
\begin{eqnarray}
\sigma \left( r_{i}\right) &=&\left\vert \frac{\varepsilon _{0}}{h_{r}\left(
r_{i}\right) ~\left[ \frac{1}{r_{j}}-\frac{1}{r_{i}}\right] \left(
-r_{i}^{2}\right) }\right\vert \left( \phi _{i}-\phi _{j}\right) =\frac{r_{j}%
}{r_{i}}\frac{\varepsilon _{0}\left( \phi _{i}-\phi _{j}\right) }{\left\vert
r_{i}-r_{j}\right\vert }  \notag \\
\sigma \left( a\right) &=&\frac{b}{a}\frac{\varepsilon _{0}\left( \phi
_{a}-\phi _{b}\right) }{\left( a-b\right) }\ ;\ \ \sigma \left( b\right) =%
\frac{a}{b}\frac{\varepsilon _{0}\left( \phi _{b}-\phi _{a}\right) }{\left(
a-b\right) }  \label{concentric spheres S}
\end{eqnarray}%
integrating these densities on $S_{a}$ and $S_{b}$ we obtain the total
charge on such surfaces. Further, taking into account Eqs. (\ref{concentric
spheres C}, \ref{concentric spheres E}) we obtain the consistency of all
these results%
\begin{equation*}
Q_{a}=-Q_{b}=\frac{4\pi \varepsilon _{0}ab\left( \phi _{a}-\phi _{b}\right) 
}{\left( a-b\right) }=C_{11}\left( \phi _{a}-\phi _{b}\right) \ ;\ \ \mathbf{%
E}\left( \mathbf{r}\right) =\frac{Q_{b}}{4\pi \varepsilon _{0}r^{2}}\mathbf{e%
}_{r}
\end{equation*}

\subsection{Electric fields and charge densities for a prolate spheroidal
shell}

For a prolate spheroidal shell, we obtain the electric field within the
shell and the charge density on the ellipsoidal surfaces by substituting
Eqs. (\ref{prolate coordh}, \ref{GH prolate}) and (\ref{Z prolate}) in Eqs. (%
\ref{concentric spheres E}, \ref{concentric spheres S})%
\begin{eqnarray*}
\mathbf{E}\left( \xi ,\eta ,\varphi \right) &=&\frac{\phi _{i}-\phi _{j}}{a%
\sqrt{\sinh ^{2}\xi +\sin ^{2}\eta }~\left[ \ln \left( \tanh \frac{\xi _{j}}{%
2}\right) -\ln \left( \tanh \frac{\xi _{i}}{2}\right) \right] ~\sinh \xi }%
\mathbf{e}_{\xi }\left( \xi ,\eta ,\varphi \right) \\
\sigma \left( u_{i},\eta ,\varphi \right) &=&\left\vert \frac{\varepsilon
_{0}}{a\sqrt{\sinh ^{2}\xi _{i}+\sin ^{2}\eta }~\left[ \ln \left( \tanh 
\frac{\xi _{j}}{2}\right) -\ln \left( \tanh \frac{\xi _{i}}{2}\right) \right]
~\sinh \xi _{i}}\right\vert \left( \phi _{i}-\phi _{j}\right) \ \ ;\ \
i,j=1,2\ \ \text{and\ \ }i\neq j
\end{eqnarray*}%
the magnitude of the electric field and the surface charge densities are
independent of the azimutahl angle $\varphi $ as the symmetry suggests.
Calculation of electric fields and surface charge densities in other complex
geometries is also straightforward.

\section{Chains of embedded conductors\label{sec:embedding}}

\begin{figure}[tbh]
\begin{center}
{\small \includegraphics[width=7.2cm]{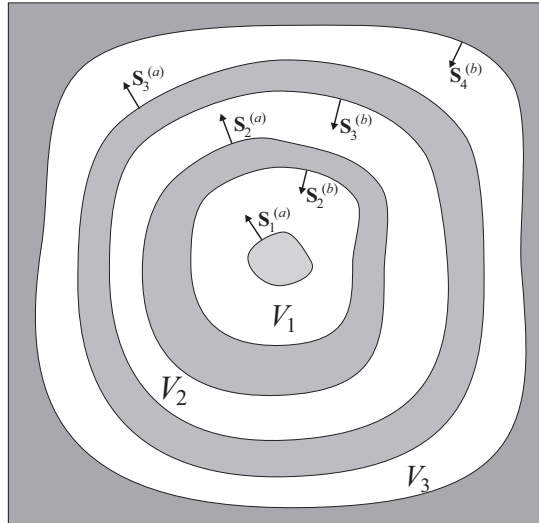} }
\end{center}
\caption{Configuration of succesive embedding of 4 conductors. The surface $%
S_{1}^{\left( a\right) }$ has only outer part while the surface $%
S_{4}^{\left( b\right) }$ has only an inner part (the surface of the
cavity). Conductors 2 and 3 have an inner part $S_{i}^{\left( b\right) }\ $%
and an outer part $S_{i}^{\left( a\right) }$. }
\label{fig:embedding}
\end{figure}

Let us assume that we have a set of $N+1\ $conductors succesively embedded
as indicated in Fig. \ref{fig:embedding}. We label them as$\ 1,2,\ldots ,N+1$
from the inner to the outer. We can see that for $k=2,\ldots ,N$ the surface 
$S_{k}$ of the $k-th$ conductor has an inner part that we denote as $%
S_{k}^{\left( b\right) }$, and an outer part that we denote as $%
S_{k}^{\left( a\right) }$. However, for $S_{1}$ we only have outer part%
\footnote{%
Let us recall that even if the most interior conductor has a cavity, we can
exclude its surface for calculations.}, while for $S_{N+1}$ we only have an
inner part. Nevertheless, we still preserve the inner and outer notation for 
$S_{1}$ and $S_{N+1}\ $by writing $S_{1}=S_{1}^{\left( a\right) }$ and $%
S_{N+1}=S_{N+1}^{\left( b\right) }$. The volume $V_{k}$ with $k=1,\ldots ,N$
is defined between $S_{k}^{\left( a\right) }$ and $S_{k+1}^{\left( b\right)
} $.

A set of conductors in succesive embedding has some particular properties
(see appendix C in Ref \cite{capac2}). First, the only non-zero coefficients
of capacitance are the diagonal ones $C_{ii}\ $and the terms of the form $%
C_{i,i\pm 1}$. Of course, for $i=1$ there is no term of the form $C_{i,i-1}$%
, and for $i=N+1$ there is no term of the form $C_{i,i+1}$. The diagonal
terms are related with the non-diagonal ones as follows%
\begin{eqnarray}
C_{ii} &=&-\left( C_{i,i-1}+C_{i,i+1}\right) \ \ \ ;\ \ \ i=2,\ldots ,N 
\notag \\
C_{11} &=&-C_{12}\ \ ;\ \ C_{N+1,N+1}=-C_{N,N+1}  \label{cembedded}
\end{eqnarray}%
now, owing to the symmetry of the matrix and the properties (\ref{cembedded}%
), it is enough to calculate the non-diagonal terms of the form $C_{i,i+1}$
for $i=1,\ldots ,N$. It can also be proved that \cite{capac2}%
\begin{equation}
C_{i,i+1}=-\varepsilon _{0}\oint_{S_{i}^{\left( a\right) }}\nabla
f_{i+1}\left( u_{i}^{\left( a\right) }\right) \cdot \mathbf{n}_{i}^{\left(
a\right) }dS\ \ ;\ \ i=1,\ldots ,N  \label{capac embedding}
\end{equation}%
that is, only the outer part of the surface $S_{i}$ contributes to the
surface integral in the calculation of$\ C_{i,i+1}$. It worths pointing out
that the properties described above are valid for any geometry of the
conductors, the only condition is that they should be in succesive embedding.

Now let us add the hypotheses we have been working with. That is, that we
have a curvilinear coordinate system $\left( u,v,w\right) $ such that the
coordinate $u$ acquires a constant value $u_{p}^{\left( \mathcal{\lambda }%
\right) }$ on each surface $S_{p}^{\left( \lambda \right) }$ with $%
p=1,2,\ldots ,N+1$ and $\lambda =a,b$.

First of all, we want to calculate the BHF's within each volume $V_{i}$. By
definition of the BHF's, they satisfy the boundary conditions%
\begin{equation}
f_{i}\left( S_{i}^{\left( \lambda \right) }\right) =1,\ \ f_{i}\left(
S_{j}^{\left( \lambda \right) }\right) =0\ \ ;\ \ i\neq j\text{ \ and\ \ }%
\lambda =a,b  \label{BHF condic}
\end{equation}%
It could be shown that within the volumen $V_{i}$ the BHF's have the
following properties%
\begin{equation*}
f_{i}\left( V_{i}\right) +f_{i+1}\left( V_{i}\right) =1\ \ ;\ \
f_{i+k+1}\left( V_{i}\right) =f_{i-k}\left( V_{i}\right) =0\text{\ };\ \
i=1,2,\ldots ,N\ \ ,\ \ k\geq 1
\end{equation*}%
therefore, it is sufficient to calculate $f_{i+1}$ within the volume $V_{i}$%
. Our boundary conditions for the BHF $f_{i+1}$\ within $V_{i}$ are given by%
\begin{equation*}
f_{i+1}\left( S_{i}^{\left( a\right) }\right) =0\ \ \ ;\ \ f_{i+1}\left(
S_{i+1}^{\left( b\right) }\right) =1
\end{equation*}%
with the same analysis done in section \ref{Laplace}, we see that the
functions $f_{i+1}$ only depends on $u$, such that these boundary conditions
can be written as\footnote{%
Note in passing that the fact that only $f_{i}$ and $f_{i+1}$ are non-zero
in $V_{i}$ could be understood with the boundary conditions on $V_{i}$. For
instance, the boundary conditions on $V_{i}$ for $f_{i-1}$ would be $%
f_{i-1}\left( u_{i}^{\left( a\right) }\right) =0$ and $f_{i-1}\left(
u_{i+1}^{\left( b\right) }\right) =0$, by uniqueness the only solution
within $V_{i}$ is $f_{i-1}\left( V_{i}\right) =0$.}%
\begin{equation}
f_{i+1}\left( u_{i}^{\left( a\right) }\right) =0\ \ \ ;\ \ f_{i+1}\left(
u_{i+1}^{\left( b\right) }\right) =1  \label{boundary fi+1}
\end{equation}%
now, taking into account that in this context the volume $V_{i}$ in which we
are evaluating the BHF's is delimited by the two surfaces $S_{i}^{\left(
a\right) }$ and $S_{i+1}^{\left( b\right) }$, we can use the same arguments
that led to Eqs. (\ref{regla de oro}) with the assignments 
\begin{equation}
i\rightarrow i\ \ ,\ j\rightarrow i+1,\ u_{i}\rightarrow u_{i}^{\left(
a\right) }\ ,\ \ u_{j}\rightarrow u_{i+1}^{\left( b\right) }
\label{assignments}
\end{equation}
From these facts the function $f_{i+1}$ evaluated at any point within the
volume $V_{i}\ $yields%
\begin{equation}
f_{i+1}\left( V_{i}\right) =\frac{Z\left( u\right) -Z\left( u_{i}^{\left(
a\right) }\right) }{Z\left( u_{i+1}^{\left( b\right) }\right) -Z\left(
u_{i}^{\left( a\right) }\right) }\ \ ;\ \ i=1,\ldots ,N
\label{regla de oro3}
\end{equation}%
With a similar procedure we can show that%
\begin{eqnarray}
f_{i+1}\left( V_{i+1}\right) &=&\frac{Z\left( u\right) -Z\left(
u_{i+2}^{\left( b\right) }\right) }{Z\left( u_{i+1}^{\left( a\right)
}\right) -Z\left( u_{i+2}^{\left( b\right) }\right) }\ \ ;\ \ i=1,\ldots ,N-1
\label{regla de oro4} \\
f_{i+1}\left( V_{i+1+k}\right) &=&f_{i+1}\left( V_{i-k}\right) =0\ \ ;\ \
k\geq 1  \label{regla de oro5}
\end{eqnarray}%
On the other hand, the coefficients of capacitance can be obtained from (\ref%
{capac embedding}) and \ref{regla de oro3}, with the same arguments that led
from (\ref{eq:CapacitanciaDiazHerrera}) to (\ref{eq:Capacitancia_Nueva}).
Thus taking into account the assignments (\ref{assignments}) and the fact
that only the outer part of the surface $S_{i}$ contributes to the surface
integral in the calculation of$\ C_{i,i+1}$, we obtain%
\begin{equation}
C_{i,i+1}=-\frac{\varepsilon _{0}}{\left\vert Z\left( u_{i+1}^{\left(
b\right) }\right) -Z\left( u_{i}^{\left( a\right) }\right) \right\vert }%
\oint_{S_{i}^{\left( a\right) }}\left\vert H\left( v,w\right) \right\vert
dvdw\ ;\ \ i=1,2,\ldots ,N  \label{eq:Capacitancia_Nueva2}
\end{equation}

\subsection{Succesive embedding of concentric spheres}

A simple example consists of examining the succesive embedding of $N+1$
concentric spherical conductors. In that case $u\equiv r$ and we define a
set of radii in the form 
\begin{equation}
a_{1}<b_{2}<a_{2}<b_{3}<a_{3}<\ldots <b_{N}<a_{N}<b_{N+1}
\label{concentric radii}
\end{equation}%
note that $b_{i}$ refers to the radius of the cavity of the $i-th$ conductor
while $a_{i}$ refers to the radius of the external surface of the $i-th$
conductor. Hence, $a_{i}-b_{i}$ is the thick of the $i-th$ conductor.

Using (\ref{regla de oro3}) and (\ref{Z(u)spheres}), we obtain the solution
for $f_{i+1}$ in a volume $V_{i}$ (i.e. between the radii $a_{i}$ and $%
b_{i+1}$)%
\begin{eqnarray}
f_{i+1}\left( r\right) &=&\frac{Z\left( r\right) -Z\left( r_{i}^{\left(
a\right) }\right) }{Z\left( r_{i+1}^{\left( b\right) }\right) -Z\left(
r_{i}^{\left( a\right) }\right) }=\frac{Z\left( r\right) -Z\left(
a_{i}\right) }{Z\left( b_{i+1}\right) -Z\left( a_{i}\right) }=\frac{\frac{1}{%
r}-\frac{1}{a_{i}}}{\frac{1}{b_{i+1}}-\frac{1}{a_{i}}}  \notag \\
f_{i+1}\left( r\right) &=&\frac{a_{i}b_{i+1}}{a_{i}-b_{i+1}}\left( \frac{1}{r%
}-\frac{1}{a_{i}}\right) \ \ \ \text{for\ \ }a_{i}\leq r\leq b_{i+1}
\label{BHFsphere}
\end{eqnarray}%
and substituting Eqs. (\ref{Z(u)spheres}, \ref{Grsolminus})\ in Eq. (\ref%
{eq:Capacitancia_Nueva2})\ the coefficient of capacitance $C_{i,i+1}$ yields

\begin{eqnarray}
C_{i,i+1} &=&-\frac{\varepsilon _{0}\oint_{S_{i}^{\left( a\right)
}}\left\vert H\left( \theta ,\varphi \right) \right\vert d\theta ~d\varphi }{%
\left\vert Z\left( r_{i+1}^{\left( b\right) }\right) -Z\left( r_{i}^{\left(
a\right) }\right) \right\vert }=-\frac{\varepsilon _{0}\oint_{S_{i}^{\left(
a\right) }}\sin \theta ~d\theta ~d\varphi }{\left\vert Z\left(
b_{i+1}\right) -Z\left( a_{i}\right) \right\vert }=-\frac{4\pi \varepsilon
_{0}}{\left\vert \frac{1}{b_{i+1}}-\frac{1}{a_{i}}\right\vert }  \notag \\
C_{i,i+1} &=&-\frac{4\pi \varepsilon _{0}a_{i}b_{i+1}}{b_{i+1}-a_{i}}
\label{capacsphere}
\end{eqnarray}

\section{Curvilinear coordinates and geometrical factors\label{ap:coordinate
geom}}

We have obtained coefficients of capacitance in terms of the generalized
curvilinear coordinates $\left( u,v,w\right) $ that are adapted to their
symmetries. Notwithstanding, since such coefficients are purely geometrical
in nature, it is more useful to obtain formulas for them in terms of
geometrical factors (radii, major semiaxes, focal distances, etc.) instead
of generalized coordinates $\left( u,v,w\right) $. Hence, in this appendix
we rewrite the curvilinear functions obtained for the capacitances in terms
of geometrical factors.

\subsection{Prolate confocal spheroidal shell}

Recalling the equation of an ellipsoid of revolution (revolved around the $%
Z- $axis) in cartesian coordinates, we have%
\begin{equation}
\frac{z^{2}}{a_{i}^{2}}+\frac{x^{2}+y^{2}}{b_{i}^{2}}=1
\label{ec elipsoid pro}
\end{equation}%
for a prolate spheroid, $a_{i}$ refers to the major semiaxis and $b_{i}$ to
the minor semiaxis of the ellipsoid of revolution generated by $\xi _{i}$.
Comparing equations (\ref{hiper id pro1}, \ref{ec elipsoid pro}) and taking
into account that $\cosh \xi $ and $\cosh \xi $ are positive for $\xi >0$ we
have%
\begin{equation}
\cosh \xi =\frac{a_{i}}{a}\ \ \text{and \ }\sinh \xi =\frac{b_{i}}{a}
\label{sicta geompro}
\end{equation}%
we shall also take into account that%
\begin{equation}
a_{i}^{2}-b_{i}^{2}=a^{2}  \label{aibiarelated}
\end{equation}%
by using a hyperbolic identity, and applying Eqs. (\ref{sicta geompro}) we
have%
\begin{eqnarray}
\tanh \frac{\xi _{i}}{2} &=&\frac{\sinh \xi _{i}}{1+\cosh \xi _{i}}=\frac{%
b_{i}/a}{1+\frac{a_{i}}{a}}=\frac{b_{i}}{a+a_{i}}  \notag \\
\tanh \frac{\xi _{i}}{2} &=&\frac{\sqrt{a_{i}^{2}-a^{2}}}{a+a_{i}}
\label{tanhsictaigeom}
\end{eqnarray}%
note that the focal distance $d=2a$ is the same for all ellipses by
definition. Therefore, it is convenient to put the expressions in terms of
the major semiaxes $a_{i},\ a_{j}\ $of each ellipsoid and the (common) focal
semi-distance $a$. Substituting (\ref{tanhsictaigeom}) in Eq. (\ref%
{capacprolate}) we obtain the coefficients of capacitance for a prolate
confocal spheroidal shell 
\begin{equation}
C_{ii}=-C_{ij}=\frac{4\pi a\varepsilon _{0}}{\left\vert \ln \left( \frac{%
\left( a+a_{i}\right) \sqrt{a_{j}^{2}-a^{2}}}{\left( a+a_{j}\right) \sqrt{%
a_{i}^{2}-a^{2}}}\right) \right\vert }  \label{capacprolate3}
\end{equation}%
it is interesting to check that despite equation (\ref{capacprolate3}) is
not defined at $a=0$, the limit $a\rightarrow 0$ exists and reduces to the
case of concentric spheres as it must be. Such a limit can be found by using
L'Hopital's rule. Assuming $a_{j}>a_{i}$ we have%
\begin{eqnarray*}
\lim_{a\rightarrow 0}\frac{4\pi \varepsilon _{0}}{C_{ii}} &=&\lim_{a%
\rightarrow 0}\frac{\ln \left[ \left( a+a_{i}\right) \sqrt{a_{j}^{2}-a^{2}}%
\right] -\ln \left[ \left( a+a_{j}\right) \sqrt{a_{i}^{2}-a^{2}}\right] }{a}
\\
&=&\lim_{a\rightarrow 0}\left[ \frac{\left( \sqrt{a_{j}^{2}-a^{2}}-\frac{%
\left( a+a_{i}\right) a}{\sqrt{a_{j}^{2}-a^{2}}}\right) }{\left(
a+a_{i}\right) \sqrt{a_{j}^{2}-a^{2}}}-\frac{\left( \sqrt{a_{i}^{2}-a^{2}}-%
\frac{\left( a+a_{j}\right) a}{\sqrt{a_{i}^{2}-a^{2}}}\right) }{\left(
a+a_{j}\right) \sqrt{a_{i}^{2}-a^{2}}}\right] \\
&=&\lim_{a\rightarrow 0}\left[ \frac{\left( a_{i}-a\right) \left(
a_{j}^{2}-2a^{2}-a_{i}a\right) -\left( a_{j}-a\right) \left(
a_{i}^{2}-2a^{2}-a_{j}a\right) }{\left( a_{i}^{2}-a^{2}\right) \left(
a_{j}^{2}-a^{2}\right) }\right] \\
\lim_{a\rightarrow 0}\frac{4\pi \varepsilon _{0}}{C_{ii}} &=&\frac{\left(
a_{j}-a_{i}\right) }{a_{i}a_{j}}
\end{eqnarray*}%
by comparing with (\ref{concentric spheres C}) we see that the limit of
concentric spheres is obtained appropriately.

\subsection{Oblate confocal spheroidal shell}

In the case of an oblate ellipsoid of revolution (revolved around the $Z-$%
axis), its equation in cartesian coordinates, yield%
\begin{equation}
\frac{z^{2}}{b_{i}^{2}}+\frac{x^{2}+y^{2}}{a_{i}^{2}}=1
\label{ec elipsoid obl}
\end{equation}%
comparing Eqs. (\ref{ec elipsoid obl}, \ref{ec elipsoid obl2}) and using Eq.
(\ref{aibiarelated}) we have%
\begin{eqnarray}
a^{2}\sinh ^{2}\xi _{i} &=&b_{i}^{2}\ ;\ \ a^{2}\cosh ^{2}\xi
_{i}=a_{i}^{2}\ \Rightarrow \ \ \tanh \xi _{i}=\frac{b_{i}}{a_{i}}=\frac{%
\sqrt{a_{i}^{2}-a^{2}}}{a_{i}}  \notag \\
\tanh \xi _{i} &=&\sqrt{1-\frac{a^{2}}{a_{i}^{2}}}=\sqrt{1-\epsilon _{i}^{2}}
\label{tanhsic}
\end{eqnarray}%
where we have taken into account that $a/a_{i}=\epsilon _{i}$, where\ $%
\epsilon _{i}$ denotes the eccentricity of the ellipse associated with the
coordinate $\xi _{i}$. Substituting (\ref{tanhsic}) in (\ref{oblategeo}) we
have

\begin{equation*}
C_{ii}=\frac{4\pi a\varepsilon _{0}}{\left\vert \sin ^{-1}\left( \sqrt{%
1-\epsilon _{j}^{2}}\right) -\sin ^{-1}\left( \sqrt{1-\epsilon _{i}^{2}}%
\right) \right\vert }
\end{equation*}%
using the identity $\cos ^{-1}x=\sin ^{-1}\sqrt{1-x^{2}}$ for $0\leq x\leq 1$
we get

\begin{equation}
C_{ii}=\frac{4\pi a\varepsilon _{0}}{\left\vert \cos ^{-1}\epsilon _{j}-\cos
^{-1}\epsilon _{i}\right\vert }  \label{oblategeom2}
\end{equation}%
which coincides with (\ref{oblategeom}). Once again, the limit with $%
a\rightarrow 0$ reduces correctly to the case of concentric spheres.

\subsection{Two confocal elliptical cylinders}

By assuming that the major semiaxis $a_{i}$ goes along $X$ and the minor
semiaxes goes along $Y$ we have%
\begin{equation*}
\frac{x^{2}}{a_{i}^{2}}+\frac{y^{2}}{b_{i}^{2}}=1
\end{equation*}%
and comparing with (\ref{elipcyl3}), we obtain%
\begin{equation}
\tanh \xi _{i}=\frac{b_{i}}{a_{i}}  \label{tanhsicta}
\end{equation}%
we shall use the identity%
\begin{equation}
\tanh ^{-1}x=\frac{1}{2}\ln \left( \frac{1+x}{1-x}\right) \ \ ;\ \
\left\vert x\right\vert <1  \label{id tanhinv}
\end{equation}%
by applying (\ref{id tanhinv}) and (\ref{aibiarelated}) in Eq. (\ref%
{tanhsicta}) yields%
\begin{eqnarray*}
\xi _{i} &=&\tanh ^{-1}\left( \frac{b_{i}}{a_{i}}\right) =\frac{1}{2}\ln
\left( \frac{1+\frac{b_{i}}{a_{i}}}{1-\frac{b_{i}}{a_{i}}}\right) =\frac{1}{2%
}\ln \left( \frac{a_{i}+b_{i}}{a_{i}-b_{i}}\right) =\frac{1}{2}\ln \left[ 
\frac{\left( a_{i}+b_{i}\right) ^{2}}{a_{i}^{2}-b_{i}^{2}}\right] \\
\xi _{i} &=&\frac{1}{2}\ln \left[ \frac{\left( a_{i}+b_{i}\right) ^{2}}{a^{2}%
}\right] =\ln \left( \frac{a_{i}+b_{i}}{a}\right)
\end{eqnarray*}%
so that%
\begin{equation}
\xi _{i}-\xi _{j}=\ln \left( \frac{a_{i}+b_{i}}{a_{j}+b_{j}}\right) =\ln %
\left[ \frac{a_{i}+\sqrt{a_{i}^{2}-a^{2}}}{a_{j}+\sqrt{a_{j}^{2}-a^{2}}}%
\right]  \label{sictai-j}
\end{equation}%
substituting (\ref{sictai-j}) in (\ref{cijcylindellip1}) we obtain Eq. (\ref%
{cijcylindellip2}). This result is clearly independent on the assumption
that the major semiaxes goes along $X$.

\subsection{Two eccentric cylinders}

We shall find the coordinate quantity $\tau _{2}-\tau _{1}$ in terms of
geometrical factors: that is in terms of the radii $r_{1}$ and $r_{2}\ $of
the cylinders and the separation $D\ $between their centers. We shall work
in the two dimensional space with coordinates $\left( \tau ,\sigma \right) $%
. We shall start from the identities%
\begin{eqnarray}
x^{2}+\left( y-a\cot \sigma \right) ^{2} &=&\frac{a^{2}}{\sin ^{2}\sigma } 
\notag \\
y^{2}+\left( x-a\coth \tau \right) ^{2} &=&\frac{a^{2}}{\sinh ^{2}\tau }
\label{identibipolar}
\end{eqnarray}%
and the equation in cartesian coordinates for a circle of radius $r\ $%
centered at $\left( x_{c},0\right) $%
\begin{equation}
\left( x-x_{c}\right) ^{2}+y^{2}=r^{2}  \label{identibipolar1}
\end{equation}
comparing (\ref{identibipolar}) with (\ref{identibipolar1}) it can be shown
that%
\begin{equation}
a=r\sinh \tau \ \ ;\ \ x_{c}=r\cosh \tau  \label{identibipolar2}
\end{equation}%
from Eqs. (\ref{identibipolar2}) we have%
\begin{eqnarray}
a &=&r_{1}\sinh \tau _{1}=r_{2}\sinh \tau _{2}
\label{eq:Distancia entre Centros-0} \\
D &=&x_{c_{2}}-x_{c_{1}}=r_{2}\cosh \tau _{2}-r_{1}\cosh \tau _{1}
\label{eq:Distancia entre Centros-1}
\end{eqnarray}%
squaring Eq. (\ref{eq:Distancia entre Centros-1}) we have 
\begin{equation}
D^{2}=r_{2}^{2}\cosh ^{2}\tau _{2}+r_{1}^{2}\cosh ^{2}\tau
_{1}-2r_{1}r_{2}\cosh \tau _{1}\cosh \tau _{2}
\label{eq:Distancia entre Centros-1b}
\end{equation}%
and using the hyperbolic identity $\cosh ^{2}\tau -\sinh ^{2}\tau =1$, as
well as Eq. (\ref{eq:Distancia entre Centros-0}), then Eq. (\ref%
{eq:Distancia entre Centros-1b}) becomes 
\begin{eqnarray*}
D^{2} &=&r_{2}^{2}+r_{1}^{2}+r_{2}^{2}\sinh ^{2}\tau _{2}+r_{1}^{2}\sinh
^{2}\tau _{1}-2r_{1}r_{2}\cosh \tau _{1}\cosh \tau
_{2}=r_{2}^{2}+r_{1}^{2}+2\left( r_{1}\sinh \tau _{1}\right) \left(
r_{2}\sinh \tau _{2}\right) -2r_{1}r_{2}\cosh \tau _{1}\cosh \tau _{2} \\
&=&r_{2}^{2}+r_{1}^{2}+2r_{1}r_{2}\left( \sinh \tau _{2}\sinh \tau
_{1}-\cosh \tau _{1}\cosh \tau _{2}\right) \\
D^{2} &=&r_{2}^{2}+r_{1}^{2}-2r_{1}r_{2}\cosh \left( \tau _{2}-\tau
_{1}\right)
\end{eqnarray*}%
from which we obtain the desired relation%
\begin{equation}
\tau _{2}-\tau _{1}=\cosh ^{-1}\left( \frac{r_{1}^{2}+r_{2}^{2}-D^{2}}{%
2r_{1}r_{2}}\right)  \label{taosgeom}
\end{equation}

\end{document}